\documentclass[12pt]{article}
\usepackage{amsmath}
\usepackage{amssymb}
\usepackage{amsfonts}
\usepackage{amscd}
\usepackage{delarray}
\usepackage{graphicx}

\usepackage{amsmath,amsmath,amsfonts,amssymb,color,bbm}

\usepackage{graphicx}

\def\Symp#1,#2,#3,#4.{\left[\left(\begin{array}{c}#1\\#2\end{array}\right),\left(\begin{array}{c}#3\\#4\end{array}\right)\right]}
\def\Vec#1,#2.{\left(\!\begin{array}{c}#1\\#2\end{array}\!\right)}
\def\vec#1,#2.{{#1\choose{#2}}}
\def\ket#1.{|#1\rangle}

\def\bra#1.{\langle#1|}
\def\braket#1,#2.{\langle#1|#2\rangle}
\newcommand{\beq}{\begin{equation}}
\newcommand{\eeq}{\end{equation}}
\newcommand{\beqa}{\begin{eqnarray}}
\newcommand{\eeqa}{\end{eqnarray}}

\usepackage{soul}
\definecolor{purplerep}{rgb}{1,0.1,1}

\definecolor{green}{rgb}{0.2,0.6,0.8}

\begin{document}

\title{Falsification of a Time Operator Model
based on Charge-Conjugation-Parity violation of neutral K-mesons}
\date{}
\author{}
\maketitle

\centerline{Thomas Durt\footnote{Aix Marseille Universit\'e, CNRS, Centrale Marseille, Institut Fresnel, UMR 7249, 13013 Marseille, France; e-mail: thomas.durt@centrale-marseille.fr},
Antonio Di Domenico\footnote{
Department of Physics, Sapienza University of Rome, and
INFN Sezione di Roma, P.le A.~Moro, 2, I-00185 Rome, Italy; e-mail: antonio.didomenico@roma1.infn.it} and
Beatrix C. Hiesmayr\footnote{University of Vienna, Faculty of Physics, Boltzmanngasse 5, 1090 Vienna; Beatrix.Hiesmayr@univie.ac.at} }

\begin{abstract}

The founding fathers of the quantum theory already struggled with the different
roles of space and time in quantum theory. Position is by default represented by an operator, whereas time is usually treated as a parameter. Time operator models exist, in particular a certain Time Operator model denoted the Temporal Wave Function (T.W.F.). This one extends the Born's rule to the time domain. The statistics of pure exponential decay does not allow to exclude this Born rule for time. However the tiny violation of the combined discrete symmetries $P$ (parity) and $C$ (charge conjugation) observed in the K-meson system allows us, as we prove in this paper, to discriminate between the standard predictions, made in the framework of the so-called mass-decay matrix formalism, and the T.W.F. predictions. We also present experimental data of a particular $CP$ violating decay channel, that is in contrast with the prediction
of T.W.F. approach model, resulting in its falsification.

\end{abstract}

\section*{Introduction.}

In the present paper, we investigate fundamental aspects of time, which is still a controversial issue in the quantum theory (see for example Ref.~\cite{booktime}).

Our main goal is to discriminate the Temporal Wave Function (T.W.F.) approach \cite{09temporal,arxiv,entangledDurt}
from the standard approach applied to decaying meson-systems. The T.W.F. approach is in the last resort motivated by the idea that, similar to the complementary observables assigned to the position and momentum of a particle in the non-relativistic formulation of the quantum theory, there would exist an operator representing time which would be complementary to the Hamiltonian.
A Time Operator would be the temporal component of a relativistic quadri-vector containing as spatial components the observables associated with the position of the particle. As is well known in non-relativistic quantum mechanics, each spatial component is complementary relatively to the corresponding component of the momentum.
We expect thus that the Time Operator would be complementary to the Hamiltonian,  so that ultimately the quadrivector time-{\bf position} would be complementary to the quadrivector energy-{\bf momentum}.
The temporal degree of freedom would no longer be an external, classical, parameter in this approach, contrarily to the standard approach. In this view the time of occurence of an event, for instance of the decay of an unstable quantum particle, would become a stochastic variable, distributed according to a generalized Born rule, in analogy with the position of a quantum particle in non-relativistic quantum mechanics \cite{09temporal,arxiv,entangledDurt,TSO,champ}.

%{\bf General comment: do we really need to emphasize here  the role of entanglement?
%In the end we use single and not enangled kaons to disprove TWF approach!! maybe it is better to introduce only single kaons.}
%Entangled
%{\bf neutral}
%kaon pairs are appropriate candidates for applying these ideas because due to entanglement their decay processes exhibit strong temporal correlations, a property which distinguishes them from their quantum optical counterparts and even from any other source of entangled systems. Besides, pairs of entangled kaons are sent along opposite directions at the same speed, so that the distance between the source and the decay events can be used as a clock. Moreover, due to EPR correlations one of the decaying kaons can be use to herald the state of the other \st{photon} {\bf kaon}, which also provides very precise information about the time at which a given preparation occurs.

Our main result is the following: a precise measurement of the asymmetry of the decay rates of
%the asymmetry of
%neutral K-mesons initially prepared in two strangeness eigenstates,
the two possible strangeness eigenstates $K^0$ and $\overline K^0$ into pion pairs, $A_{2\pi}(t)$, makes it possible to falsify a Time Operator approach, the T.W.F. approach.

The paper is structured as follows.

In the first section, we motivate the Time Operator approach, summarizing (section \ref{1.1}) debates around the status of time that can be traced back to the founding fathers of the quantum theory (Bohr, von Neumann, Dirac, Heisenberg Schr\"odinger and Pauli). We also explain briefly why exponential decay does not make it possible to end the debate (section \ref{1.2}) \cite{champ}. We also generalize the reasoning to the most general situation of the exponential decay treated as a multi-channel process (section \ref{1.3}).

In section \ref{single}, we briefly recall the phenomenology of unstable
neutral
kaon particles and  their description in terms of the mass-decay matrix in the standard Weisskopf-Wigner
approximation (WWA) \cite{WWA,loy,chsud}.

In section \ref{proposal1} we confront the standard WWA formulation of decay processes to the Time Operator approach.  We show how a certain model of Time Operator, denoted the T.W.F. approach, can agree with the standard
%approximately fit the standard
WWA predictions in the neutral kaon system -- including the description of kaon oscillations in 
semi-leptonic decay-- provided we neglect the violation of the combined parity ($P$) and charge conjugation ($C$) symmetries (from now on $CP$ violations).
% $t\sim \tau_S$, with $\tau_S$ the $K_S$ lifetime.
Moreover, we show that the precise measurement of the $A_{2\pi}(t)$ asymmetry
at short  times performed by the CPLEAR experiment \cite{Angelopoulos,CPLEAR} -- in perfect agreement with the WWA description --
does not confirm the
prediction of our T.W.F. model, which results therefore
falsified.

The last section is devoted to discussions and conclusions of the falsification.

Extra-material can be found in appendix, which contains a detailed treatment of the Fermi golden rule when the decay is a multi-channel process.
%In summary, our goal is to discriminate between the standard description of the decay process and a non-standard approach where it is postulated that the probability density function assigned to the statistical distribution of decay times is equal to the modulus squared of a complex wave function, that we shall from now on call the Temporal Wave Function (denoted T.W.F.), and is intimately related to the Time Operator approach.  % We show that a simple version of the T.W.F. approach is fasifiable experimentally due to CP violating contributions to the decay process of kaons, although a more sophisticated, ad hoc version of the T.W.F. approach perfecly mimicks the predictions made in the framework of the (standard) mass-decay matrix formalism.

 \section{Standard and non-standard conceptions of time in quantum mechanics.}
 \subsection{Time in quantum mechanics: is there a problem?\label{1.1}}

 As intensively discussed in the literature, see e.g. Ref.~\cite{confusion}, the status of time in the quantum theory has been in the past, and is still to some extent today, the source of much confusion. Let us summarize the discussions of   \cite{confusion}, inviting the reader to consult this very detailed reference for more precise explanations and bibliography.

 The
main source of confusion could be formulated through the question:

 {\it...Is time a classical variable (c-number)?...},

 by which we mean that time can be treated as an external parameter (similar to time as it appears in classical, Newtonian, and/or relativistic, Minkowski space-time).

The other alternative can be formulated as follows:

 {\it ...Is time a quantum quantity (q-number) represented by an operator?...},

by which we mean that it is represented by or associated with an internal parameter (example: phase) of some quantum system, represented by an operator (similar to position components $\hat x, \hat y$ and $\hat z$ in the non-relativistic quantum theory).

As noted in \cite{confusion}, various answers were given by the founding fathers of the quantum theory to this deep foundational issue. Bohr, for instance, treated  time as a c-number and concluded that there is no problem in doing so, which is close to the mainstream view of contemporary physicists on the subject. Von Neumann acknowledged that time is treated as a c-number in the first quantized, non-relativistic, quantum theory but at the same time he considered that this is a source of  problems, mainly for the following reasons:

-The coordinates x, y, z are described by operators (in the non-relativistic quantum theory), and Lorentz transformation treats space and time on the same footing (see \cite{vaccaro} for a fresh view on this subject) so that time ought to be treated as a q-number, and, consequently, represented by an operator. Let us e.g. reconsider the usual quantization rule that associates the energy $E$ with $i\hbar{\partial\over \partial t}$,  $p_x$ to $\hbar{\partial\over i\partial x}$,  $p_y$ to $\hbar{\partial\over i\partial y}$, and $p_z$ to $\hbar{\partial\over i\partial z}$. It obviously possesses a strong relativistic flavour, which suggests that time ought to be represented by an operator conjugate to the Hamiltonian.

Dirac also felt strongly concerned by this question and, as noted in \cite{confusion}, he wrote his famous equation in order to formulate a Lorentz covariant quantum theory (of the electron), where space and time would be treated on the same footing. Finally, Pauli remarked that when an Hamiltonian possesses a continuous bounded spectrum, it is not possible to construct an operator $T$ such that $[\hat H,\hat T]=i\hbar \hat{\mathbb I}$. In other words, it is not always possible to treat a quantum system as a clock.

As it was noted by J. Hilgevoord \cite{confusion}, many discussions about the status of time focused on Pauli's former objection  that was since then interpreted as a no-go theorem about the existence of a Time Operator. In the meanwhile, the question of the existence of a Time Operator somewhat lost interest \cite{booktime}, and we can claim without risk that the main opinion today is that the time that appears in Schr\"odinger equation IS a classical variable (c-number), and can thus be considered as an external parameter (universal or general time). Several theoretical arguments invite us indeed to orient the debate to this conclusion, and it is commonly accepted nowadays that in quantum field theory time and even space \cite{Newton} correspond to classical parameters (the spacetime ``arena''). Despite of these theoretical arguments, some confusion remains concerning these questions \cite{booktime} and there were several attempts to associate with time an operator \cite{booktime,bauer} or a positive-operator valued measure (POVM) \cite{Kijowski,moyer} (or Superoperator  \cite{courbette,MPC2}) in order to deal with arrival times and/or decaying processes\footnote{Decay processes are appealing candidates for a Time Operator approach because to the contrary of stable systems, the probability of decay by unit of time (probability density function) of such systems is normalised to unity.}
(photon in a cavity, particle tunneling from a trap, decaying radio-active particle and so on).
\subsection{Two approaches to exponential decay.\label{1.2}}
 In the non-standard approach (Time Operator and/or Superoperator approaches), the time of decay of an individual quantum unstable system is considered to be a quantum quantity (q-number), an internal parameter, represented by an operator or a Superoperator. The Time Super Operator approach of Misra, Prigogine and Courbage {\it et al.} \cite{courbette,MPC2}), makes it for instance possible to associate a generalized Time Operator (Super Operator) with any Hamiltonian provided its spectrum is not bounded by above. It has been applied to kaon phenomenology in the past \cite{TSO}. In this paper we shall consider a simpler model, which fits into the Time Operator approach, through which we associate with the statistical distribution of decay processes a temporal wave function (T.W.F. \cite{09temporal,arxiv,entangledDurt,TSO,champ}). This model is for instance very natural and straightforward in the case of exponential decay \cite{arxiv}, in which case it is common, following Gamow, to associate with the unstable system a wave function of the form $\Psi_S(t)=\Psi_S(0)exp^{-i({mc^2\over \hbar}-i{\Gamma\over 2})\cdot t}$ which is interpreted as the amplitude of a complex energy state.

According to the standard interpretation, $ {|\Psi_S(t)|^2\over   |\Psi_S(0)|^2}$ is interpreted to be equal to the survival probability $P_s(t)$ up to time $t$ ($0\leq t$).

 Alternatively, let us define the Temporal Wave Function $\tilde \Psi^{T.W.F.}(t)$ through\footnote{From now on the upperly tilded quantities will always refer to quantities derived in the framework of the T.W.F. approach.}

 \begin{equation}\tilde \Psi^{T.W.F.}(t)=\tilde \Psi^{T.W.F.}(0)exp^{-i({mc^2\over \hbar}-i{\Gamma\over 2})\cdot t},\end{equation} with $\tilde \Psi^{T.W.F.}(0)=\sqrt \Gamma$. It is straightforward to check that \begin{equation}{-d P_s(t)\over dt}=|\tilde \Psi^{T.W.F.}(t)|^2.\end{equation} The choice $\tilde \Psi^{T.W.F.}(0)=\sqrt \Gamma$ also ensures that the survival probability decreases from its initial value (1) at time $t=0$ to its final value (0), when $t$ goes to $+\infty$.

 As we see, by a formal renormalisation, we can define a temporal wave function $\tilde \Psi^{T.W.F.}(t)$ (T.W.F.), normalised to unity over time in the interval $[0,\infty]$, such that the probability of decay by unit of time (probability density function) is equal to its modulus squared, in full analogy with the usual Schr\"odinger wave function in position representation. Considered so, the T.W.F. provides a time representation assigned to the distribution of decay times of an unstable quantum system.

It is worth noting that this is not only a formal trick, but that this procedure resists to experimental falsification in the case of purely exponential decay. Indeed, lifetimes of particles are often measured indirectly in particle physics, by fitting the energy distribution of decay products with a Breit-Wigner (Lorentzian) distribution. Also in this case the standard and Time Operator approaches cannot be distinguished \cite{champ}. Roughly summarized, the argument goes as follows:

-Suppose that the Hamiltonian is time-independent and that  at time
$t=0$
the amplitude of probability that the particle energy is $E$ equals $\psi^E(E)$. Then, developing the wave function in the energy eigenbasis it is straightforward to show that $P_s(t)=|<\psi(0),\psi(t)>|^2=|\int_{E_{min.}}^{E_{max.}}dE e^{-i{Et\over \hbar}}|\psi^E(E)|^2|^2$, where the spectrum of the Hamiltonian runs from $E_{min.}$ to $E_{max.}$. When the spectrum of the Hamiltonian is large enough $|\int_{E_{min.}}^{E_{max.}}dE e^{-i{Et\over \hbar}}|\psi^E(E)|^2|$ is close to $|\int_{-\infty}^{+\infty}dE e^{-i{Et\over \hbar}}|\psi^E(E)|^2|$ and the survival probability is equal to the squared modulus of the Fourier transform of the energy distribution. Now, the characteristic function (Fourier transform) of the Breit-Wigner distribution is equal to
$ C \cdot e^{-\mathrm{i}(mt-\frac{\mathrm{i}}{2}\Gamma |t|) }$.

Thus fitting
 with a Breit-Wigner distribution is consistent with the ``standard'' recipe: $P_s(t)=|<\psi(0),\psi(t)>|^2$  (disregarding negative times).

-Assume that the decay probability density function is associated with a ``temporal wave function'': ${-d P_s(t)\over dt}=\Gamma\theta(t)e^{-\Gamma t}$=$|\tilde \Psi^{T.W.F.}(t)|^2$ with $|\tilde \Psi^{T.W.F.}(t)|^2$=$|\sqrt{\Gamma}\theta(t)e^{-\mathrm{i}(m-\frac{\mathrm{i}}{2}\Gamma)t ) }|^2$ (where $\theta(t)$ is the Heaviside function). Then, $|\tilde \Psi^{T.W.F.}(t)|^2$=$|\int_{-\infty}^{+\infty}dE e^{+i{Et\over \hbar}}\hat{\tilde \Psi}^{T.W.F.}(E)|^2$,

  with $\hat{\tilde \psi}^{T.W.F.}(E)$=${1\over 2\pi}\int_{-\infty}^{+\infty}dt e^{-i{Et\over \hbar}}\tilde \Psi^{T.W.F.}(t)$=$-i\sqrt{{\Gamma\over 2\pi}}{1\over (E-(m-\frac{\mathrm{i}}{2}\Gamma))} $. The modulus squared of $\hat{\tilde \psi}^{T.W.F.}(E)$ is precisely equal to the Breit-Wigner distribution:

$|\hat{\tilde \psi}^{T.W.F.}(E)|^2$=$N.{1\over (E-m)^2+(\Gamma/2)^2}$ with $N$ a normalisation factor.  To fit with a Breit-Wigner distribution is thus also consistent with the ``non-standard'' recipe: ${-d P_s(t)\over dt}=\Gamma\theta(t)e^{-\Gamma t}$=$|\tilde \Psi^{T.W.F.}(t)|^2$.

 It is worth noting that in the previous paragraph we performed two different Fourier transforms: in one case we transformed amplitudes, in the other case we transformed the modulus square of these amplitudes. By chance, in the case of exponential decay, both approaches lead to the same prediction. In order to study more in depth the two approaches, it is thus necessary to investigate more complex decay processes which exhibit a non-exponential statistics. This motivates our choice to consider neutral kaons which show also subtle non trivial effects due to CP violation.

 In order to tackle these more sophisticated decay processes, we shall have to generalize the T.W.F. model described here in the case of exponential decay. This model will be outlined in section \ref{T.W.F.section}, on the basis of kaon phenomenology that we outline in section \ref{single}.

Before doing that we will show in the next paragraph how the T.W.F. approach can be applied in the case where several decay channels are present, a situation that has not been investigated in previous literature. In particular we show how the summation rule of the partial decay widths $\Gamma_i$ assigned to various decay channels can be accomodated in the framework of the T.W.F. approach.

\subsection{Multi-channel exponential decay processes.\label{1.3}}
\subsubsection{Multi-mode exponential decay processes in the standard approach.}In the standard approach, a $\Gamma$ factor is associated with each outgoing mode (decay product) through the Fermi golden rule. As
it
is shown in appendix, in the case of multi-channel exponential
decay processes, each individual decay mode contributes to the decay of the survival probability by a factor which can be shown, at the first order of perturbation theory, to  increase proportionally to $|G_{i,\lambda}|^2t^2\sin\!\mathrm{c}^2\left( \left(\omega_{i,\lambda}-\omega_{\mathrm{in}}\right) \frac{t}{2}\right),$ where $G_{i,\lambda}$ represents the coupling constant with the mode labelled by $\lambda$ of the $i$th decay species. Summing over internal degrees of freedom $\lambda$ of different decay products $i$ and also over the energies $\omega$, which is taken into account through the introduction of the D.O.S. (density of states) factor aimed at converting the discrete sum into an integral, via the prescription $D.O.S.(\omega_{in},i,\lambda)\delta\omega$=number of modes labelled by $i$ and $\lambda$ with an energy comprised between $\hbar(\omega_{in}-\delta \omega/2)$ and $\hbar(\omega_{in}+\delta \omega/2)$, and making use of the property \cite{englert}
\begin{equation} \label{dirac}
\lim_{t\rightarrow+\infty} (t \ \mathrm{sinc}^2\left[\delta\omega\frac{t}{2}\right])
 =2\pi \delta^{Dirac}(\delta \omega).
 \end{equation}
We find, after integrating over a continuum of frequencies centered around $\omega_{in}$,

\begin{eqnarray} \label{eq:WeAreDone}
\sum_{i,\lambda,\omega}| c_{i,\lambda}\left(t\right) |^2
%& &\nonumber\\
%\nonumber
&=&\lim_{t\rightarrow+\infty}\sum_{i,\lambda}\int_{-\infty}^{+\infty}\hspace{-12.5pt}\mathrm{d}\omega D.O.S.(\omega,i,\lambda) |G_{i,\lambda}(\omega)|^2 t^2\sin\!\mathrm{c}^2\!\!\left[\left(\omega_{\mathrm{in}}-\omega\right)\frac{t}{2}\right]\nonumber \\ \nonumber
    &=&... \\
   &=&\sum_{i}\Gamma_it =\Gamma t.
\end{eqnarray}
We observe that $\sum_{i,\lambda,\omega}| c_{i,\lambda}\left(t\right) |^2$ increases linearily with time. In other words the amplitude that a decay occurs increases with $\sqrt t$ which is the essence of the Fermi Golden Rule.

The important lesson that must be retained from (\ref{eq:WeAreDone}) is that the resulting $\Gamma$ factor is the sum of the contributions of all individual modes (integrated over species and internal degrees of freedom). In particular, $\Gamma$ =$\sum_i\Gamma_i$ where $i$ labels the different species of decay products (for instance
for the decay process of a kaon distinct values of $i$ are assigned e.g. to $\pi^+\pi^-$, $\pi^0\pi^0$, $\pi\ell \nu$, $3\pi^0$, $\pi^+\pi^-\pi^0$, ... decay products.
$\Gamma_i$ in turn obeys $\Gamma_i=\sum_\lambda D.O.S.(\omega_{in},i,\lambda) 2\pi|G_{i,\lambda}(\omega_{in})|^2$, where $\lambda$ does label the internal degrees of freedom of the species under concern, in agreement with the Fermi Golden Rule.

%$\Gamma$ is proportional to the sum of the squared moduli of the single mode coupling factors $G_{i,\lambda}$, so that

\subsubsection{Multi-channel exponential decay processes in the T.W.F. approach.\label{multichannel}}

As it
is shown in appendix, the multi-mode exponential decay process does not fundamentally differ from a single mode process, being given that only one outgoing mode ($|eff.>$) is effectively coupled to the ingoing mode. Of course, this effective mode is in general a coherent superposition of modes of different species, which will never interfere in practice, so that in practice we can as well consider that the sum over the different possible outgoing channels is incoherent.

Then, when the Fermi golden rule applies, it is straightforward to interpret a generation of decay products labelled by the index $i$ at a rate $\Gamma_i$ dt either in the standard approach where this rate is equal to $|\sqrt \Gamma_i\sqrt{dt}|^2$ in agreement with Fermi Golden Rule, or in the T.W.F. approach where this rate is seen to be equal to $|\sqrt \Gamma_i|^2$ $dt$. The so-called sum rule for the $\Gamma_i$ factors is seen here to derive from (\ref{eq:WeAreDone}).

   \section{Single kaons and CP-violation in the standard approach.\label{single}}
\subsection{Short overview of kaon phenomenology.}
%{\it GENERAL COMMENT: I THINK THIS PARAGRAPH IS TOO DIDACTICT FOR A SCIENTIFIC PAPER.
%SOMETHING SHOULD BE CUT.  I HAVE REWRITTEN IT.
%MOREOVER I THINK TO INTRODUCE THE KS AND KL
%STATES BEFORE THE MASS DECAY MATRIX IS CONFUSNG. I WOULD REWRITE BOTH PARAGRAPHS
%INTO A SINGLE ONE.

Neutral Kaons are bosons
that were discovered in the forties during the study of cosmic rays.
They are constituted by a quark-antiquark pair,
$\mathrm{K}^{0}$ $(=d\bar{s})$ with strangeness $S=+1$, and
$\overline{\mathrm{K}}^{0}$ $(=\bar{d}s)$ with strangeness $S=-1$.
They are usually produced in processes dominated by strong interactions,
%(which conserve strangeness)
while weak interactions -- which do not
conserve strangeness -- are responsible for their decay or induce oscillations through
transitions with an intermediate virtual state of the type $\mathrm{K}^{0} \leftrightarrow \pi\pi \leftrightarrow \overline{\mathrm{K}}^{0}$.
\par
In absence of
$CP$-violation,
the  following combinations
\begin{equation}\label{kk1}
|\mathrm{K}_1\rangle=\frac{1}{\sqrt{2}}\big{(}|\mathrm{K}^0\rangle
-|\overline{\mathrm{K}}^0\rangle\big{)},~~~
|\mathrm{K}_2\rangle=\frac{1}{\sqrt{2}}\big{(}|\mathrm{K}^0\rangle
+|\overline{\mathrm{K}}^0\rangle\big{)},
\end{equation}
which are eigenstates of the $CP$ operator, are also eigenstates of the total Hamiltonian of the system.
\par
%$CP$-\emph{violation}  was discovered by Christenson et al.
%\cite{christ}.
%$CP$-violation means that the long-lived kaon can
%also decay to ``$2\pi"$. Then,
%As it was discovered by
In 1964 Christenson et al.
discovered \cite{christ} that the $CP$ symmetry is slightly violated (by a factor of $\mathcal{O}(10^{-3}$)~)
in the neutral kaon system,
%by weak interactions,
so that the
$CP$ eigenstates $\mathrm{K}_1$ and $\mathrm{K}_2$ do not exactly coincide
with the
%are not
%exact
eigenstates of the Hamiltonian.
In fact these contain a small $CP$ impurity
and can be
expressed as coherent superpositions of the $\mathrm{K}_1$ and
$\mathrm{K}_2$ states\footnote{At this level, CPT invariance will be assumed according to the CPT theorem \cite{Lueders}.}:
\begin{eqnarray}\label{kk2}
|\mathrm{K}_S\rangle=\frac{1}{\sqrt{1+|\epsilon|^2}}\big{[}
|\mathrm{K}_1\rangle +\epsilon ~ |\mathrm{K}_2\rangle \big{]},\\
|\mathrm{K}_L\rangle=\frac{1}{\sqrt{1+|\epsilon|^2}}\big{[}
\epsilon
~|\mathrm{K}_1\rangle + |\mathrm{K}_2\rangle \big{]},\label{kk3}
\end{eqnarray}
where $\epsilon$ is a complex $CP$-violation parameter \cite{Olive},
%$|\epsilon|\ll1$

\begin{equation}\label{expepsilon}
|\epsilon| = ( 2.228 \pm 0.011)\times10^{-3}, ~~~
\mathrm{arg}(\epsilon) = (43.5 \pm 0.5)^\circ.
\end{equation}
$\mathrm{K}_{S}$ state and
$\mathrm{K}_{L}$ are the short- and long-lived states, with quite different lifetimes,
%The lifetimes of the $\mathrm{K}_{S}$ state and
%$\mathrm{K}_{L}$ are rather different
$\tau_{S}\approx 8.92\times10^{-11}\mathrm{s}$ and $\tau_{L}\approx 5.17\times10^{-8}\mathrm{s}$, respectively.

%Neglecting
Direct $CP$ violation in the decay is negligible \cite{Olive,directcp}, therefore
%the weak disintegration process
one can distinguish
the $\mathrm{K}_{1}$ state, which cannot decay into $3\pi^0$, a pure $CP=-1$ final state,
from the $\mathrm{K}_{2}$ one, which cannot decay into $\pi\pi$ ($\pi^+\pi^-$ or $\pi^0\pi^0$), a pure $CP=+1$ final state.
Moreover, the sign of the lepton (or of the pion) in semileptonic decays, $\pi^-\ell^+\nu$ or $\pi^+\ell^-\bar{\nu}$, exclusively identifies a $\mathrm{K}^{0}$ or a
$\overline{\mathrm{K}}^{0}$ state, respectively, assuming the validity of the $\Delta S =\Delta Q$ rule.
%
% ``$2\pi$" and not into ``$3\pi$'',while the
%$\mathrm{K}_{2}$ states can decay into ``$3\pi$" and not into ``$2\pi$''

%by semileptonic decays, $\pi^-\ell^+\nu$ and $\pi^+\ell^-\bar{\nu}$ which exclusively identify $\mathrm{K}^{0}$ and
%$\overline{\mathrm{K}}^{0}$, respectively, thanks to the $\Delta S =\Delta Q$ rule}.
%\st{
%$\mathrm{K}_1$
%and $\mathrm{K}_2$ are the decay modes of kaons. }
%
%In absence of
%$CP$-violation,
%{\bf
%$\mathrm{K}_1$
%and $\mathrm{K}_2$ are the physical decaying states;
%}
%the weak disintegration process distinguishes the
%$\mathrm{K}_{1}$ states which can decay into ``$2\pi$" and not into ``$3\pi$'',while the
%$\mathrm{K}_{2}$ states can decay into ``$3\pi$" and not into ``$2\pi$''

\par
Inverting equations (\ref{kk2},\ref{kk3}) in the basis $|\mathrm{K}^0\rangle$ and
$|\overline{\mathrm{K}}^0\rangle$,  we get
\begin{eqnarray}\label{kk02}
|\mathrm{K}_0\rangle=\frac{\sqrt{1+|\epsilon|^2}}{\sqrt{2}(1+\epsilon)}\big{[}
|\mathrm{K}_S\rangle +|\mathrm{K}_L\rangle \big{]},\\
|\overline{\mathrm{K}}^0\rangle=\frac{\sqrt{1+|\epsilon|^2}}{\sqrt{2}(1-\epsilon)}\big{[}
|\mathrm{K}_S\rangle -|\mathrm{K}_L\rangle \big{]}.\label{kk03}
\end{eqnarray}

\subsection{Mass-decay matrix.}
In the WWA mass-decay formalism, it is assumed that the evolution law describing the surviving kaon can be cast in Schr\"odinger form \cite{WWA}
\begin{equation}i\hbar{\partial\over \partial t}|\psi(t)\rangle=H|\psi(t)\rangle,\label{schrod}
\end{equation} where $H$, the
effective, not hermitian, Hamiltonian, can be written as a mass-decay matrix \cite{loy,chsud}
which has
 the following form (for instance in the basis $|\mathrm{K}^0\rangle$ and
$|\overline{\mathrm{K}}^0\rangle$, taking here $\hbar=c=1$):
\begin{equation}
H=M-\frac{\mathrm{i}}{2}\Gamma\equiv
\begin{array}({cc}) M_{11}-\frac{\mathrm{i}}{2}\Gamma_{11} &
M_{12}-\frac{\mathrm{i}}{2}\Gamma_{12}  \\
M_{21}-\frac{\mathrm{i}}{2}\Gamma_{21} &
M_{22}-\frac{\mathrm{i}}{2}\Gamma_{22}
\end{array}\label{5}
\end{equation}
where $M$ and $\Gamma$ are
individually Hermitian. $\Gamma$ incorporates contributions from all possible decay channels. For instance, the decay of a  $|\mathrm{K}_1\rangle$ ($|\mathrm{K}_2\rangle$) state into
{\bf $\pi\pi$ ($3\pi^0$)} contributes to the mass-decay matrix by a factor $\Gamma_{K_1\rightarrow \pi\pi}|\mathrm{K}_1\rangle\langle \mathrm{K}_1|$ ($\Gamma_{K_2\rightarrow \pi^0\pi^0\pi^0}|\mathrm{K}_2\rangle\langle \mathrm{K}_2|$). In principle all
$\Gamma$ factors associated with the various decay channels can be computed thanks to the Fermi Golden Rule.

The $\Gamma$ matrix can then be expressed by summing the contributions from all possible decay channels:
\begin{equation}\Gamma=\sum_i\Gamma_{|\mathrm{K}_i\rangle\rightarrow i}|\mathrm{K}_i\rangle\langle \mathrm{K}_i|\label{decaym}\end{equation}
For instance, we find, writing explicitly some of the decay products:
\begin{eqnarray}\Gamma  = & & \sum_i\Gamma_{|\mathrm{K}_i\rangle\rightarrow i}|\mathrm{K}_i\rangle\langle \mathrm{K}_i|
\nonumber\\
  \approx &  &
 \Gamma_{|\mathrm{K}_1\rangle\rightarrow 2\pi}|\mathrm{K}_1\rangle\langle \mathrm{K}_1|\label{decaym}
 +
  \Gamma_{|\mathrm{K}_2\rangle\rightarrow 3\pi}|\mathrm{K}_2\rangle\langle \mathrm{K}_2|\label{decaym}
 \nonumber\\
 & + &
 \Gamma_{|\mathrm{K}^0\rangle\rightarrow \pi^-\ell^+\nu}|\mathrm{K}^0\rangle\langle \mathrm{K}^0|\label{decaym}
 +
 \Gamma_{|\mathrm{\bar{K}}^0\rangle\rightarrow \pi^+\ell^-\bar{\nu}}|\mathrm{\bar{K}}^0\rangle\langle \mathrm{\bar{K}}^0|
 \nonumber\\
 &  +& ...
 \nonumber
\end{eqnarray}

If for instance we have to compute the decay rate that a prepared $K_2$ state decays into a $\ell^+$ through a semi-leptonic decay process, we find:
$$\Gamma_{|\mathrm{K}_2\rangle\rightarrow \pi^-\ell^+\nu}=\langle K_2 | \Gamma | K_2 \rangle_{\pi^-\ell^+\nu} =  \Gamma_{|\mathrm{K}^0\rangle\rightarrow \pi^-\ell^+\nu} \cdot
 \langle K_2 |  \mathrm{K}^0\rangle\langle \mathrm{K}^0| K_2 \rangle$$

%{\it SEMILEPTONIC DECAYS ARE NOT MEDIATED BY STRONG INTERACTIONS. DO WE NEED THE FOLLOWING SENTENCE?}\green{REPLY: NO BUT I WOULD LIKE TO CLARIFY SOME QUESTIONS HERE...}
%\st{So called semi-leptonic decay processes for instance are mediated by strong interactions, so that they distinguish $|\mathrm{K}^0\rangle$ and $|\mathrm{\overline K}^0\rangle$ states. For the $|\mathrm{K}^0\rangle$-labelling ($|\mathrm{\overline K}^0\rangle$-labelling) semi-leptonic decay processes, $|\mathrm{K}_i\rangle=|\mathrm{K}^0\rangle (|\mathrm{\overline K}^0\rangle$).}

$\mathrm{K}_L$ and $\mathrm{K}_S$ are the eigenstates of $H$.
The eigenvalues of the mass-decay
matrix corresponding to the long and short states are equal to
\begin{equation}\label{mas}
E_L=m_L-\frac{\mathrm{i}}{2}\Gamma_L,~~~\mathrm{ and} ~~~
E_S=m_S-\frac{\mathrm{i}}{2}\Gamma_S.
\end{equation}
Reintroducing dimensional factors we get
\begin{equation}\label{mas}
E_L=m_Lc^2-\frac{\mathrm{i}}{2}\hbar\Gamma_L,~~~\mathrm{ and} ~~~
E_S=m_Sc^2-\frac{\mathrm{i}}{2}\hbar\Gamma_S.
\end{equation}
%\st{Experimental bounds established that $\frac{(m_L-m_S)c^2}{-(\Gamma_L-\Gamma_S)\hbar}\approx \triangle m c^2
%\tau_S/\hbar\approx 0.5$}

%\st{The $CP$-violation was established by the observation that
%$\mathrm{K}_L$ decays not only via three-pion, which has natural
%$CP$ parity, but also via the two-pion (``$2\pi$") modes with an
%experimentally observed violation amplitude of the order of $10^{-3}$. }

These features are taken into account by noting that when a $|\mathrm{K}^0\rangle$ state is prepared at time $t=0$, the state of the kaon at time $t$ is, in virtue of equations (\ref{schrod},\ref{kk2},\ref{kk3},\ref{kk02}), equal to

%\begin{equation}(1/\sqrt 2)(exp^{-iE_St/\hbar}|\mathrm{K}_S\rangle+\epsilon exp^{-iE_Lt/\hbar}|\mathrm{K}_L\rangle+(1/\sqrt 2)(exp^{-iE_Lt/\hbar}|\mathrm{K}_L\rangle+\epsilon exp^{-iE_St/\hbar}|\mathrm{K}_S\rangle
%\end{equation}

\begin{eqnarray} |\mathrm{K}^0(t)\rangle=\frac{\sqrt{1+|\epsilon|^2}}{\sqrt{2}(1+\epsilon)}(exp^{-iE_St/\hbar}|\mathrm{K}_S\rangle+exp^{-iE_Lt/\hbar}|\mathrm{K}_L\rangle)\label{kkk}
\end{eqnarray}
% ={(exp^{-iE_St/\hbar}+\epsilon exp^{-iE_Lt/\hbar})|\mathrm{K}_1\rangle+(exp^{-iE_Lt/\hbar}+\epsilon exp^{-iE_St/\hbar})|\mathrm{K}_2\rangle)\over \sqrt 2 (1+\epsilon) }
Accordingly, in the
WWA
standard approach, the probability per unit of time that at time $t$ the kaon particle decays say in the CP=+1 sector by emission of a pair of pions can be computed by estimating the corresponding contribution to the decay matrix which is, in virtue of (\ref{decaym}), equal to
%\st{$\Gamma_{K_1\rightarrow \pi\pi}\cdot \langle  \mathrm{K}^0(t)|\mathrm{K}_1\rangle\langle \mathrm{K}_1|\mathrm{K}^0(t)\rangle$}
$\Gamma_{K_1\rightarrow \pi\pi}\cdot \left| \langle \mathrm{K}_1|\mathrm{K}^0(t)\rangle \right|^2$
where $\Gamma_{K_1\rightarrow \pi\pi}$ is the probability of transition
per unit of time (or production rate) between a $K_1$ state and a pair of pions.

Thus we find:
\begin{equation}P_{K^0(0)\rightarrow 2\pi(t)}={\Gamma_{K_1\rightarrow \pi\pi}\over 2|1+\epsilon|^2}|exp^{-iE_St/\hbar}+\epsilon exp^{-iE_Lt/\hbar}|^2,\label{singlepairrate}\end{equation}
%This can be computed by estimating the corresponding contribution to the decay matrix which is, in virtue of (\ref{decaym}) $\Gamma_{K_1\rightarrow \pi\pi}\langle  \mathrm{K}^0(t)|\mathrm{K}_1\rangle\langle \mathrm{K}_1|\mathrm{K}^0(t)\rangle$l,
where $2\pi(t)$ denotes the production rate at time $t$ of a pair of pions (for instance $2\pi=\pi^0\pi^0$ , $\pi^+\pi^-$).

%By similar computations we find, making use of the relations $\Gamma_{  \mathrm{K}_L\rightarrow 2\pi}=\Gamma_{  \mathrm{K}_1\rightarrow 2\pi}\cdot \langle  \mathrm{K}_L|\mathrm{K}_1\rangle\langle \mathrm{K}_1|\mathrm{K}_L\rangle$ and $\Gamma_{  \mathrm{K}_S\rightarrow 2\pi}=\Gamma_{  \mathrm{K}_1\rightarrow 2\pi}\cdot \langle  \mathrm{K}_S|\mathrm{K}_1\rangle\langle \mathrm{K}_1|\mathrm{K}_S\rangle$.

%In the standard approach, we also find $|\epsilon|^2=\Gamma_{  \mathrm{K}_L\rightarrow 2\pi}/\Gamma_{   \mathrm{K}_S\rightarrow 2\pi}$,  a relation that can be found in the literature.
Anagously for the $CP=-1$ decay into $3\pi^0$ we get:

%In what follows, we shall most often limit ourselves to CP labelling decay products, for instance, $\pi^0\pi^0$ and/or $\pi^+\pi^-$, or $3\pi^0$, for which we respectively find, beside the already established relation} (\ref{singlepairrate}) \st{the relation}

\begin{equation}P_{K^0(0)\rightarrow 3\pi(t)}=  {\Gamma_{K_2\rightarrow \pi^0\pi^0\pi^0}\over 2|1+\epsilon|^2}|exp^{-iE_Lt/\hbar}+\epsilon exp^{-iE_St/\hbar}|^2\label{rate-}.\end{equation}

The production rates of all possible species of decay products can be in principle computed in a similar fashion.

%\st{where the indices $K^0(0)\rightarrow 2\pi$ (resp. $K^0(0)\rightarrow 3\pi$) denotes decay processes with a specific CP=+1(resp.-1) decay product.}

\section{Experimental proposal: falsifying the Time Operator approach in presence of CP violation related effects.\label{proposal1}}\subsection{Time Operator Approach: a binary model.\label{T.W.F.section}}
Let us now stick to the Time Operator approach, in order to investigate whether or not the statistical distribution of decay times exhibited by
%\st{an entangled kaon pair}
 a neutral kaon
might be derived from a ``temporal wave function''. This T.W.F. would constitute a generalisation in the time domain of Born's interpretation of the quantum wave function as representing an amplitude of probability of ``spatial position''. According to the Born rule, the probability that at a given time a particle is located around a given position is supposed to  be proportional to the modulus squared of a complex field, the so-called quantum wave function.
In the temporal wave function approach the probability that a ``click'' occurs at a given time in a given detector is also supposed to  be linked to a complex field, the Temporal Wave Function, and we shall propose in the next section an experiment aimed at challenging/falsifying this hypothesis. Before we do so, let us develop a Time Operator formalism aimed at modeling neutral kaon phenomenology. Our model which has been developed elsewhere \cite{09temporal,arxiv,entangledDurt} is directly inspired from the spin 1/2 formalism, in particular from the non-relativistic spinorial wave function approach developed by Pauli. In this approach, the wave function of a spin 1/2 particle splits into two components, the spin ``up'' component $\Psi_+(x,y,z,t)$ and the spin ``down'' component $\Psi_-(x,y,z,t)$, relatively to an a priori defined axis of reference. Both components depend on space and time, and according to the standard interpretation, the probability that at time $t$ and location $(x,y,z)$ a measurement of the spin projection along the axis of reference (traditionally denoted the $z$ axis) provides the result spin up (down) is equal to $|\Psi_+(x,y,z,t)|^2/(|\Psi_+(x,y,z,t)|^2+|\Psi_-(x,y,z,t)|^2)$ ($|\Psi_-(x,y,z,t)|^2/(|\Psi_+(x,y,z,t)|^2+|\Psi_-(x,y,z,t)|^2)$). Besides, the probability that the particle is present in a neighbourhood $dxdydz$ around $(x,y,z)$ is equal to ($|\Psi_+(x,y,z,t)|^2+|\Psi_-(x,y,z,t)|^2)dxdydz$.

The Temporal Wave Function differs from the standard spin 1/2 wave function considered above in the following sense: the binary value taken by the spin projection along an axis of reference ($\pm \hbar/2$) is in our case replaced by the CP eigenvalue ($\pm 1$), while space is replaced by time. Accordingly, instead of dealing with probabilities to observe spin up or spin down we are now considering the probabilities that a single kaon decays in the CP=+1 or -1 sectors. We assume thus in the T.W.F. approach that there exists a two-components temporal wave function $(\tilde \Psi^{T.W.F.}_+(t),\tilde \Psi^{T.W.F.}_-(t))$ such that the probability that the kaon decays by emitting a CP=+1 (-1) labelled decay product in the time interval $[t,t+dt]$ (where $t\geq 0$ and $dt>0$) is equal to $|\tilde \Psi^{T.W.F.}_+(t)|^2dt$ ($|\tilde \Psi^{T.W.F.}_-(t)|^2dt$). This binary model is somewhat reductionist because it implicitly assumes that only CP labelled decay products are emitted during the decay process of a kaon.
%{\it THIS I DO NOT UNDERSTAND. APART CP VIOLATION, A KL DECAYS
%\\- 67\% OF THE TIMES IN SEMILEPTONIC DECAYS WITH CP UNDEFINED (CP+1 or CP-1 ????)
%\\- 20\% OF THE TIMES IN $3\pi^0$
%\\ - 12\% OF THE TIMES IN $\pi^+\pi^-\pi^0$.
%\\
A priori, this is a very crude assumption, because it is known for intance that a $K_L$ decays 67 $\%$ of the times into semileptonic decays with undefined CP value, and only 20 $\%$ of the times in $3\pi^0$, and 12 $\%$ of the times into $\pi^+\pi^-\pi^0$ which both have a well-defined CP-value. 

As we shall discuss now, even semi-leptonic decay can be accomodated in the framework of the T.W.F. model, so that our reductionist hypothesis is fully justified as far as we may consistently neglect CP violation.
% {\it
%YOU ARE SAYING THAT YOU ARE NEGLECTING ABOUT 80\% OF KL DECAYS. THIS HAS NOTHING TO DO WITH CP VIOLATION.}\green{I do not mean to say this...}

{\bf Without CP violation.}
If there were no CP violation ($\epsilon=0$), then, in the standard WWA approach, the mass matrix and the decay matrix as well would both be diagonal in the CP eigen-basis ($|K_1\rangle$, $|K_2\rangle$): $|K_S\rangle=|K_1\rangle$, $|K_L\rangle=|K_2\rangle$
and only CP-labelled decay products would be emitted during the decay process, for instance CP=+1 $\pi^0\pi^0$, $\pi^+\pi^-$ states and CP=-1 $\pi^0\pi^0\pi^0$ states, but also ${1\over \sqrt 2}\cdot(|\pi^-\ell^+\nu\rangle+| \pi^+\ell^-\bar{\nu}\rangle)$ and ${1\over \sqrt 2}\cdot(|\pi^-\ell^+\nu\rangle-| \pi^+\ell^-\bar{\nu}\rangle)$, which are CP=$\pm 1$ decay products. 

In the case of the semileptonic decay this approach is not conventional, but in practice it is equivalent to the usual approach, for several reasons.
\\
The first reason is that, resulting from the fact that semileptonic decay is a CP respecting process, $$\Gamma_{|\mathrm{K}^0\rangle\rightarrow \pi^-\ell^+\nu} =  \Gamma_{|\mathrm{\bar{K}}^0\rangle\rightarrow \pi^+\ell^-\bar{\nu}}$$ so that
$$\Gamma_{|\mathrm{K}^0\rangle\rightarrow \pi^-\ell^+\nu}|\mathrm{K}^0\rangle\langle \mathrm{K}^0|
+\Gamma_{|\mathrm{\bar{K}}^0\rangle\rightarrow \pi^+\ell^-\bar{\nu}}|\mathrm{\bar{K}}^0\rangle\langle \mathrm{\bar{K}}^0| =\Gamma_{|\mathrm{K}^0\rangle\rightarrow \pi^-\ell^+\nu}(|\mathrm{K_1}\rangle\langle \mathrm{K_1}|
+|\mathrm{K_2}\rangle\langle \mathrm{K_2}|).$$
\\
The second reason is that, making use of the linearity of Schr\"odinger equation, it is equivalent to describe a decay process of the type $|\mathrm{K}^0\rangle\rightarrow
|\pi^-\ell^+\nu\rangle$ ($|\mathrm{\bar{K}}^0\rangle\rightarrow | \pi^+\ell^-\bar{\nu}\rangle$) as a symmetric (antisymmetric) fifty fifty coherent superposition of the process $|\mathrm{K_1}\rangle\rightarrow (1/\sqrt 2)(|\pi^-\ell^+\nu \rangle+ |\pi^+\ell^-\bar{\nu}\rangle)$ with the process $|\mathrm{K_2}\rangle\rightarrow (1/\sqrt 2)(|\pi^-\ell^+\nu \rangle- |\pi^+\ell^-\bar{\nu}\rangle)$.
\\
Finally, it is worth noting that detectors are not sensitive to the phase between $\pi^-\ell^+\nu$ and $\pi^+\ell^-\bar{\nu}$ decay products, which are experimentally detected in different outgoing channels, as already emphasised in the section \ref{multichannel}, where we showed that in last resort decay products are actually coherent superpositions of distinct decay species. Taken all together, these features make it possible to properly reproduce strangeness oscillations during the semileptonic decay in the framework of the T.W.F. approach which is in a sense remarkable for such a crude model.
%so that $$\mathrm{K}^0\rangle\langle \mathrm{K}^0|$$}

%\Gamma_{|\mathrm{K}^0\rangle\rightarrow \pi^-\ell^+\nu}|\mathrm{K}^0\rangle\langle \mathrm{K}^0|\label{decaym}

% \Gamma_{|\mathrm{\bar{K}}^0\rangle\rightarrow \pi^+\ell^-\bar{\nu}}|\mathrm{\bar{K}}^0\rangle\langle \mathrm{\bar{K}}^0|
%{\it AND SEMILEPTONIC DECAYS??? ARE THEY CP+ OR CP- ?? YOU CANNOT NEGLECT THEM.
%OR NOT? I THINK IT'S BETTER NOT TO MENTION THIS LABELLING EXPLICITLY.}. \rep{Here is a crucial step: my proposal is to also mention here the {\bf HOPEFULLY} CP=$\pm 1$ decay products

%{\\ \it YES BUT K2 CAN DECAY ALSO IN SEMILEPTONIC. THIS SHOUD BE TAKEN INTO ACCOUNT.
%\\
%IN OTHER WORDS, EXPERIMENTALLY:
%$$\Gamma_S \approx \Gamma_1 \approx \Gamma_{K_1\rightarrow \pi^+\pi^-}+
%\Gamma_{K_1\rightarrow \pi^0\pi^0}$$
%$$\Gamma_L \approx \Gamma_2 \gg \Gamma_{K_2\rightarrow 3\pi^0}$$
%\\
%NOW I WOULD LIKE TO PROPOSE A CHANGE OF STRATEGY.
%WHY DON'T WE GET RID OF 3PI0 DECAYS AND FOCUS ONLY CP+1 DECAYS INTO 2PI?
%IN THE END FOR THE FALSIFICATION WE ONLY NEED TO COMPARE WITH 2PI DECAYS RESULTS
%(AS I'LL SHOW LATER).
%IT WOULD HAVE SEVERAL ADVANTAGES. \green{Reply: the strategy is PARTIALLY conditioned by the result of the discussion above. But I agree with you that it is sufficient to focus on the 2$\pi$ production rate for 2 reasons: 1) it has been effectively measured with high accuracy; 2) the asymmetry paramter only depends on this quantity.}
%THEREFORE I PROPOSE TO DROP IN THE FOLLOWING ALL EXPRESSIONS WITH
%$K_2 \rightarrow 3\pi$.}

From now on, we shall only be concerned with the 2$\pi$ production rate, due to the fact that this quantity has been measured with very high accuracy. This does not mean that we fully neglect other decay channels. Actually they are still implicitly present through the normalisation which imposes the following conditions:

$\sum_{\mathrm{i=all possible CP+1 decay products}}\Gamma_{K_1\rightarrow i}=\Gamma_S=\Gamma_1$

$\sum_{\mathrm{j=all possible CP-1 decay products}}\Gamma_{K_2\rightarrow j}=\Gamma_L=\Gamma_2$

In absence of CP violation, the expression of the CP=$\pm 1$ components of the temporal wave function in terms of complex energies $E_{S(L)}$ would in turn be extremely simple:

\begin{equation}(\tilde \Psi^{T.W.F.\epsilon=0}_+(t),\tilde \Psi^{T.W.F.\epsilon=0}_-(t))=(\alpha \sqrt{\Gamma_S}exp^{-iE_St/\hbar}, \beta \sqrt{\Gamma_L}exp^{-iE_Lt/\hbar}),\label{tata}\end{equation}

%where \st{$E_{S(L)}=m_{S(L)}c^2-i(\hbar\Gamma_{S(L)}/2)$, with $m_{S(L)}$ the rest mass of the short lived (long lived) kaon state, and $\Gamma_{S(L)}$ the corresponding decay factor;} {\it (ALREADY DEFINED ABOVE)}
where $\alpha$ and $\beta$ are two complex amplitudes of which the value is fixed by the preparation of the kaon at time $t=0$.

For instance, whenever a source produces $|\mathrm{K}^0\rangle$ states at time $t=0$, which are fifty-fifty superpositions of CP=+1 and -1 states, (see e.g. equations
 (\ref{kk1},\ref{kk2},\ref{kkk},\ref{singlepairrate},\ref{rate-})), we would have $\alpha=\beta=\sqrt{1/2}$.

The tilded expressions\footnote{We remind the reader that in the following, tilded quantities are evaluated within the T.W.F. approach, while
non-tilded ones denote standard WWA expressions.}
 differ from the non-tilded ones by the normalisation factors $\sqrt{\Gamma_S}$ and $\sqrt{\Gamma_L}$:

$(\alpha,\beta)=(\Psi_+(t=0),\Psi_-(t=0))=(\tilde \Psi^{T.W.F.\epsilon=0}_+(t=0)/\sqrt{\Gamma_S},\tilde \Psi^{T.W.F.\epsilon=0}_-(t=0)/\sqrt{\Gamma_L})$.

It is easy to check that the moduli squared of the components of $(\tilde \Psi^{T.W.F.\epsilon=0}_+(t),\tilde \Psi^{T.W.F.\epsilon=0}_-(t))$ correctly reproduce \cite{arxiv} the standard predictions concerning decays in the CP=$\pm 1$ sectors (\ref{singlepairrate},\ref{rate-})  in the case $\epsilon=0$, for instance we get
\begin{equation}P_{K^0(0)\rightarrow \pi\pi(t)}^{\epsilon=0}= {\Gamma_{K_1\rightarrow \pi\pi}\over 2}
|exp^{-iE_St/\hbar}|^2\label{rate+eps0}=|\sqrt{\Gamma_{K_1\rightarrow \pi\pi}\over \Gamma_S}\tilde \Psi^{T.W.F.\epsilon=0}_+(t)|^2=\tilde P_{K^0(0)\rightarrow \pi\pi(t)}^{\epsilon=0},\end{equation}
%\begin{equation}P_{K^0(0)\rightarrow \pi^0\pi^0\pi^0(t)}^{\epsilon=0}= {\Gamma_{K_2\rightarrow \pi^0\pi^0\pi^0}\over 2}|exp^{-iE_Lt/\hbar}|^2\label{rate-eps0}=|\sqrt{\Gamma_{K_2\rightarrow \pi^0\pi^0\pi^0}\over \Gamma_L}\tilde \Psi^{T.W.F.\epsilon=0}_-(t)|^2=\tilde P_{K^0(0)\rightarrow \pi^0\pi^0\pi^0(t)}^{\epsilon=0}\end{equation}
where we dropped the $T.W.F.$ subscripts but kept the ``tilded'' notations for the production rates in the CP=+1 and -1 sectors derived in the T.W.F. approach, in order to differentiate them from the
 ``non-tilded'' predictions of the standard WWA approach (\ref{singlepairrate},\ref{rate-})).
%{\it (I WOULD AVOID TO REPEAT THIS. AT THIS POINT IT'S CLEAR WHICH IS THE WWA.)}
 We made use here of the results of section \ref{multichannel} according to which the T.W.F. amplitude of a particular species of decay products ($i$) is equal to $\sqrt{\Gamma_i}$. In the next section we shall show that in the case of departure from exponential decay due to CP violation, the standard and ``tilded'' predictions made in the framework of the binary T.W.F. model presented in the previous section differ so that a discrimination is possible.

{\bf With CP violation.}

If now we wish to incorporate CP related effects  into our T.W.F. model, we are forced to introduce a short-lived and a long-lived T.W.F.:

\begin{eqnarray}\label{kk2bis}
|\mathrm{\tilde K}_S\rangle=\frac{1}{\sqrt{1+|\tilde \epsilon_S|^2}}\big{[}
|\mathrm{K}_1\rangle +\tilde \epsilon_S ~ |\mathrm{K}_2\rangle \big{]};|\mathrm{\tilde K}_L\rangle=\frac{1}{\sqrt{1+|\tilde \epsilon_L|^2}}\big{[}
\tilde \epsilon_L
~|\mathrm{K}_1\rangle + |\mathrm{K}_2\rangle \big{]},
\end{eqnarray}
It is worth noting that we can express the short-lived and long-lived states in this way without losing generality, because the four amplitudes $\langle \mathrm{\tilde K}_S|\mathrm{ K}_1\rangle$, $\langle \mathrm{\tilde K}_S|\mathrm{ K}_2\rangle$, $\langle \mathrm{\tilde K}_L|\mathrm{ K}_1\rangle$, and $\langle \mathrm{\tilde K}_L|\mathrm{ K}_2\rangle$ obviously differ from zero. Expressed in matricial form, the corresponding pseudo-spinorial T.W.F. explicitly depend on time through

\begin{equation}(\tilde \Psi^{T.W.F.}_{+S}(t),\tilde \Psi^{T.W.F.}_{-S}(t))=\frac{1}{\sqrt{1+|\tilde \epsilon_S|^2}}\sqrt{\Gamma_S}exp^{-iE_St/\hbar}(1, \tilde \epsilon_S),\label{toto}\end{equation}

\begin{equation}(\tilde \Psi^{T.W.F.}_{+L}(t),\tilde \Psi^{T.W.F.}_{-L}(t))=\frac{1}{\sqrt{1+|\tilde \epsilon_L|^2}}\sqrt{\Gamma_L}exp^{-iE_Lt/\hbar}(\tilde \epsilon_L, 1),\label{tutu}\end{equation}

where the indices + and - refer to the CP=+1 and -1 sectors respectively. $\tilde \epsilon_S$ and $\tilde \epsilon_L$ are complex $CP$-violation parameters. According to the well-established kaon phenomenology, CP violation is small so that they are supposedly small parameters. Otherwise, we do not assume anything about their (complex) value\footnote{In previous works \cite{09temporal,entangledDurt,TSO}, we sticked to the particular choice $\tilde \epsilon_S=\tilde \epsilon_L=\tilde \epsilon$ but here we do no want to diminish the level of generality of the T.W.F. model that we consider.}.

Consequently, we shall consider in the following temporal wave functions of the form
%\footnote{Absorbing the factors  $\frac{N_S }{\sqrt{1+|\tilde \epsilon_S|^2}}$ and $\frac{N_L }{\sqrt{1+|\tilde \epsilon_L|^2}}$ into the amplitudes $\alpha$ and $\beta$.}

\begin{eqnarray}(\tilde \Psi^{T.W.F.}_+(t),\tilde \Psi^{T.W.F.}_-(t))=\nonumber \\(\alpha  \sqrt{\Gamma_S}exp^{-iE_St/\hbar}+\beta  \tilde \epsilon_L \sqrt{\Gamma_L}exp^{-iE_Lt/\hbar},\beta \sqrt{\Gamma_L}exp^{-iE_Lt/\hbar}+ \alpha \tilde \epsilon_S \sqrt{\Gamma_S}exp^{-iE_St/\hbar}),\label{bip}\end{eqnarray}

%\begin{eqnarray}(\tilde \Psi^{T.W.F.}_+(t),\tilde \Psi^{T.W.F.}_-(t))=\nonumber \\(\alpha \sqrt{\Gamma^{2\pi}_S}exp^{-iE_St/\hbar}+\beta  \tilde \epsilon_L \sqrt{\Gamma_L}exp^{-iE_Lt/\hbar},\beta N_L\sqrt{\Gamma_L}exp^{-iE_Lt/\hbar}+ \alpha N_S\tilde \epsilon_S \sqrt{\Gamma^{2\pi}_S}exp^{-iE_St/\hbar}),\label{bip}\end{eqnarray}

 where $\alpha$ and $\beta$ are fixed by the initial preparation process.
\subsection{Experimental discrimination between the WWA and T.W.F. predictions.\label{falsi}}
\subsubsection{Binary T.W.F. predictions in presence of CP-violation.\label{binaire}}
Having in mind that CP-violation is small, it is temptating to generalize our binary
T.W.F. model in presence of CP violation to a binary model where all decay products would have a clear CP signature.
%{\it LET ME ADD ANOTHER COMMENT. AS FAR AS I UNDERSTAND IN THE BINARY MODEL YOU NEED TO CLASSIFY
%DECAYS INTO TWO CLASSES: CP=+1 AND CP=-1, THAT IS THE DECAYS OF THE K1 STATE, AND THE DECAYS OF THE K2 STATE. THE FACT %THAT, EXPERIMENTALLY, THE DECAYS OF THE K1 STATE ARE EASIER TO DISTINGUISH FROM THE REST, WHILE DECAYS OF THE K2 STATE ARE %MORE DIFFICULT TO DISENTANGLE, UNLESS YOU NEGLECT 80\% OF ITS DECAYS, SHOULD NOT HAVE AN IMPACT ON THE CONSTRUCTION OF %THE MODEL. AM I RIGTHT?}
%From the beginning, this model is somewhat reductionist. It departs for instance from the standardWWA approach where
%\st{it is assumed that the mass-decay matrix also contains} CP violating contributions {\bf can be naturally accommodated in the formalism.}
%\st{(like e.g. those assigned to semi-leptonic decay processes} {\it SEMILEPTONIC DECAYS DO NOT  VIOLATE CP - FOR THE MAIN PART AT LEAST}).

%{\it(I WOULD SAY THIS TO AVOID TO STICK TO 2PI OR 3PI DECAY PRODUCTS)}
% so that the binary T.W.F. model, despite its roughness, allows us to make predictions in this case. Actually the binary T.W.F. model is also %reminiscent of the spin 1/2 phenomenology where a unique spin basis {\bf (CP basis for kaons)} is privilegged.
%\st{ , which establishes a distinction with kaons which may decay into decay products that belong to non-compatible bases (like e.g. flavor %and CP eigenbases).} %This feature does not contradict the basic principles of quantum physics however because decay occurs in one basis %at a time. Nature chooses however undeterministically  ``in which basis'' the decay will occur.

In the binary T.W.F. approach, we assume that the probability that a CP=+1 (-1) decay product is produced between time $t$ and time $t+dt$ is equal (in very good approximation) to the decrease of the CP=+1 (-1) component during that period. In particular, when a short (long) state is prepared at time $t=0$,  the decay is purely exponential and the production rates of CP=+1
$\pi\pi$
decay products
are predicted to be equal to :

%\st{
%$(1/(1+|\tilde \epsilon_S|^2))\Gamma_S exp^{-\Gamma_St}dt$ ($(|\tilde \epsilon_L|^2/(1+|\tilde \epsilon_L|^2)) \Gamma_L exp^{-\Gamma_Lt}dt$).
%}
%{\it(WE CAN COMPACT THE DISCUSSION HERE)}
%\st{
%Making use of} (\ref{rate+eps0}), \st{when a short (long) state is prepared at time $t=0$, the production rate of a $\pi\pi$ pair between time $t$ and time $t+dt$ is thus assumed to satisfy, in the binary T.W.F. model, the condition
%}
%\\
$\tilde P_{K_S(0)\rightarrow 2\pi(t)}=|\sqrt{\Gamma_{K_1\rightarrow \pi\pi}\over \Gamma_S}|^2(1/(1+|\tilde \epsilon_S|^2))\Gamma_S exp^{-\Gamma_St}dt$

($\tilde P_{K_L(0)\rightarrow 2\pi(t)}=|\sqrt{\Gamma_{K_1\rightarrow \pi\pi}\over \Gamma_S}|^2(|\tilde \epsilon_L|^2/(1+|\tilde \epsilon_L|^2)) \Gamma_L exp^{-\Gamma_Lt}dt$).

This is an assumption motivated by the fact that most often the decay process is CP-respecting, which justifies to ''import'' here an equation (\ref{rate+eps0}) valid in the $\epsilon=0$ limit. From this point of view the binary T.W.F. model is a minimal extension of the $\epsilon=0$ case\footnote{It is exactly at this level that the binary T.W.F. model departs from the WWA mass-decay formalism.}.

It is instructive to compare these prediction with the standard WWA ones. For instance, when a short (long) state is prepared at time $t=0$, the production rate of a $\pi\pi$ pair between time $t$ and time $t+dt$ is:
(\ref{schrod},\ref{kk2},\ref{kk3}),

$P_{K_S(0)\rightarrow 2\pi(t)}=|\sqrt{\Gamma_{K_1\rightarrow \pi\pi}\over \Gamma_S}|^2(1/(1+|\epsilon|^2))\Gamma_S exp^{-\Gamma_S}dt$

($P_{K_L(0)\rightarrow 2\pi(t)}=|\sqrt{\Gamma_{K_1\rightarrow \pi\pi}\over \Gamma_S}|^2(|\epsilon|^2/(1+|\epsilon|^2)) \Gamma_S exp^{-\Gamma_L}dt$).

%{\it (I WOULD SUGGEST TO DROP THE FOLLOWING SENTENCE AND FORMULAS FOR 3PI.)}
%{\\ \it =============================\\}
%Similarly, when a short (long) state is prepared at time $t=0$, the \st{production} {\bf decay} rate {\bf into } \st{of a} $\pi^0\pi^0\pi^0$ triplet is :
%\st{predicted, in the standard model, to obey}

%$P_{K_S(0)\rightarrow 3\pi(t)}=|\sqrt{\Gamma_{K_2\rightarrow \pi^0\pi^0\pi^0}\over \Gamma_L}|^2(|\epsilon|^2/(1+|\epsilon|^2))\Gamma_L exp^{-\Gamma_S}dt$

%($P_{K_L(0)\rightarrow 3\pi(t)}=|\sqrt{\Gamma_{K_2\rightarrow \pi^0\pi^0\pi^0}\over \Gamma_L}|^2(1/(1+|\epsilon_L|^2)) \Gamma_L exp^{-\Gamma_L}dt$).
%{\\ \it =============================UNTIL HERE \\}

It is already obvious at this level that the standard predictions  a priori differ from those made in the binary T.W.F. model. Considering e.g. that $\epsilon$, $\tilde \epsilon_L$ and $\tilde \epsilon_S$ are small parameters, we notice that discrepancies occur between the expression of $P_{K_L(0)\rightarrow 2\pi(t)}$ and of its tilded counterpart which is of the order of $(\sqrt{\Gamma_S}/\sqrt{\Gamma_L})^2$, which is not a small parameter.
Discarding all contributions quartic in $\epsilon,\epsilon_S$ and $\epsilon_L$ which are supposedly small parameters (see also appendix for a more detailed discussion), the contradiction is avoided at first sight provided we assume that
%\begin{equation}
%\st{$ \epsilon_L\sqrt{{\Gamma_L\over \Gamma_S}}=\epsilon=\epsilon_S\sqrt{{\Gamma_S\over \Gamma_L}}.$}
%\end{equation}
%{\it (IN THE ABOVE EXPRESSION TILDAS ARE MISSING!!! REPLACE WITH THE FOLLOWING):}
\begin{eqnarray}
\tilde \epsilon_L &=& \epsilon \sqrt{{\Gamma_S\over \Gamma_L}}
\label{approx}
\end{eqnarray}
However, it is possible in principle to discriminate experimentally the standard predictions from those derived in the binary T.W.F. approach modulo equation (\ref{approx}), as we shall show now.

\subsubsection{Incompatibility between {\bf T.W.F. and WWA} approaches.\label{modulo}}

%The initial preparation at time $t=0$ is realized through the observation of so-called semi-leptonic decay of one component of an EPR kaon pair \cite{?} (cfr appendix of ref.\cite{extended}). During this process it is possible to infer the flavor (diagonal in the $K^0$-$\overline K^0$ basis) of the surviving kaon by observing the flavor decay of the other, first decaying kaon, at time $t=0$. Then, subsequent transitions at time $t$ into a CP-labelling decay product (pion pairs in the CP=+1 sector and triplets in the CP=-1 sector) can in principle be measured accurately.

In agreement with previous notations, let us denote $P_{X(0)\rightarrow Y(t)}$ the production rate at time $t$ of a specific decay product denoted $Y$ (in what follows, $Y$most often represents a pair of pions ($\pi\pi$)), obtained after preparing a kaon in the $X$ ($X$=$K^0$ or $\overline K^0$) state at time $t=0$, evaluated in the standard WWA approach. Let us denote  $\tilde P_{X(0)\rightarrow Y(t)}$ the same rate derived in the framework of the T.W.F. approach.

The initial preparation fixes the coefficients $\alpha$ and $\beta$ in expression (\ref{bip}), making use of the identities

$(1,1)={1\over 1-\tilde \epsilon_S\tilde \epsilon_L}[(1-\tilde \epsilon_L)(1,\tilde \epsilon_S)+(1-\tilde \epsilon_S)(\tilde \epsilon_L,1)]$ and

$(1,-1)={1\over 1-\tilde \epsilon_S\tilde \epsilon_L}[(1+\tilde \epsilon_L)(1,\tilde \epsilon_S)-(1+\tilde \epsilon_S)(\tilde \epsilon_L,1)],$

which is the non-standard counterpart of the standard expressions (\ref{kk02},\ref{kk03}) which can be rewritten in a similar fashion as:

$(1,1)={1\over 1+ \epsilon}[(1,\epsilon)+( \epsilon,1)]$ and

$(1,-1)={1\over 1- \epsilon}[(1,\epsilon)-( \epsilon,1)].$

Straightforward computations similar to those of the previous section show that, in the case that a $K^0$ state is prepared at time $t=0$, the tilded quantities are:
\begin{equation}\tilde P_{K^0(0)\rightarrow 2\pi(t)}= {\Gamma_{K_1\rightarrow \pi\pi}\over 2|1-\tilde \epsilon_S\tilde \epsilon_L|^2}|(1-\tilde \epsilon_L)exp^{-iE_St/\hbar}+(1-\tilde \epsilon_S)\tilde \epsilon_L\sqrt{\Gamma_L\over  \Gamma_S} exp^{-iE_Lt/\hbar}|^2\label{nsrate+},\end{equation}

to compare with the standard WWA relation (\ref{singlepairrate}) that we reproduce below

\begin{equation}P_{K^0(0)\rightarrow 2\pi(t)}={\Gamma_{K_1\rightarrow \pi\pi}\over 2|1+\epsilon|^2}|exp^{-iE_St/\hbar}+\epsilon exp^{-iE_Lt/\hbar}|^2,\nonumber \end{equation}

In the case that a $\overline K^0$ state is prepared at time $t=0$ we get:

\begin{equation}\tilde P_{\overline K^0(0)\rightarrow 2\pi(t)}= {\Gamma_{K_1\rightarrow \pi\pi}\over 2|1-\tilde \epsilon_S\tilde \epsilon_L|^2}|(1+\tilde \epsilon_L)exp^{-iE_St/\hbar}-(1+\tilde \epsilon_S)\tilde \epsilon_L\sqrt{\Gamma_L\over  \Gamma_S} exp^{-iE_Lt/\hbar}|^2\label{overlinensrate+},\end{equation}

to compare with its standard counterpart
\begin{equation} P_{\overline K^0(0)\rightarrow 2\pi(t)}={\Gamma_{K_1\rightarrow \pi\pi}\over 2|1-\epsilon|^2}|exp^{-iE_St/\hbar}-\epsilon exp^{-iE_Lt/\hbar}|^2, \label{overlinesinglepairrate}\end{equation}

In order to discriminate the T.W.F. approach from the WWA approach on the basis of experimental data, let us now consider the asymmetry in the production rates of pion pairs, denoted $A_{\pi\pi}(t)$ and defined as follows:

\begin{equation}A_{\pi\pi}(t)= {P_{\overline K^0(0)\rightarrow 2\pi(t)}-P_{ K^0(0)\rightarrow 2\pi(t)}\over P_{ K^0(0)\rightarrow 2\pi(t)}+P_{\overline K^0(0)\rightarrow 2\pi(t)}}\end{equation}

In the standard WWA approach, making use\footnote{
We are neglecting here the tiny contribution $\mathcal{O}(10^{-6})$ to the $A_{\pi\pi}(t)$ given by direct CP violation 
\cite{Olive,directcp}.
%There even exist more sophisticated models than the standard model considered here, including CPT violation \cite{bernabeu}, for which the asymmetry exhibits the same behaviour as predicted in the standard WWA approach, but we shall not consider them in the present paper, in order not to overload our presentation.
}
of equations (\ref{singlepairrate},\ref{overlinesinglepairrate}), it is easy to show that the
 asymmetry vanishes when $t$ goes to 0: \begin{equation}\label{zero}\lim_{t\rightarrow 0}A_{\pi\pi}(t)=0\end{equation}

 %: \st{is equal to 0} \st{(at all orders in $\epsilon$). }
%{\it(I WOULD NOT STRESS THIS BECAUSE IT IS TRUE ONLY NEGLECTING DIRECT CP VIOLATION. IN ANY CASE IT %
%IS A VERY WELL KNOW RESULT AND I WOULD AVOID THE EXPLICIT COMPUTATION BELOW.)}
{%\\ \it ======== DROP ======= \\ }
%{\it ( CHANGE SIGN)}:
%\begin{eqnarray}lim_{t\rightarrow 0}A_{\pi\pi}(t)\nonumber\\=lim_{t\rightarrow 0}
%{({\Gamma_{K_1\rightarrow \pi\pi}\over 2})({1\over |1+\epsilon|^2}|exp^{-iE_St/\hbar}+\epsilon exp^{-iE_Lt/\hbar}|^2-{1\over |1-\epsilon|^2}|exp^{-iE_St/\hbar}-\epsilon exp^{-iE_Lt/\hbar}|^2)\over P_{ K^0(0)\rightarrow 2\pi(t)}+P_{\overline K^0(0)\rightarrow 2\pi(t)}}\nonumber \\= 0\label{zero}\end{eqnarray}
%{\\ \it ===================== \\ }
%\begin{eqnarray} A_{\pi\pi}(t=0) \simeq 0 \label{zero}
%\end{eqnarray}

On the contrary, if we use equations (\ref{nsrate+},\ref{overlinensrate+}) in order to evaluate the same limit in the T.W.F. approach, adopting the value $\tilde \epsilon_L=\sqrt{{\Gamma_S\over \Gamma_L}}\epsilon$ in accordance with equation (\ref{approx}), we find instead that
\begin{eqnarray}\lim_{t\rightarrow 0}\tilde A_{\pi\pi}(t)=
%lim_{t\rightarrow 0}
%{({\Gamma_{K_1\rightarrow \pi\pi}\over 2|1-\tilde \epsilon_S\tilde \epsilon_L|^2})(|(1-\tilde \epsilon_L)exp^{-iE_St/\hbar}+(1-\tilde %\epsilon_S)\tilde \epsilon_L\sqrt{\Gamma_L\over  \Gamma_S} exp^{-iE_Lt/\hbar}|^2)\over \tilde P_{ K^0(0)\rightarrow 2\pi(t)}+ \tilde %P_{\overline K^0(0)\rightarrow 2\pi(t)}}
%\nonumber \\ -lim_{t\rightarrow 0} {({\Gamma_{K_1\rightarrow \pi\pi}\over 2|1-\tilde \epsilon_S\tilde \epsilon_L|^2})(|(1+\tilde %\epsilon_L)exp^{-iE_St/\hbar}-(1+\tilde \epsilon_S)\tilde \epsilon_L\sqrt{\Gamma_L\over  \Gamma_S} exp^{-iE_Lt/\hbar}|^2)\over  \tilde %P_{ K^0(0)\rightarrow 2\pi(t)}+ \tilde P_{\overline K^0(0)\rightarrow 2\pi(t)}}
\nonumber \\\approx 2Re.((1-\sqrt{\Gamma_L\over  \Gamma_S})\cdot \tilde \epsilon_L)\approx 2Re(\tilde \epsilon_L)\approx 8 \%\not=0\label{epsilon}\end{eqnarray}

The discrepancy between the predictions (\ref{zero}) and (\ref{epsilon}) shows the possibility to experimentally discriminate between the standard and T.W.F. approaches.

 \subsubsection{Experimental falsification of the T.W.F. model.}

The transition probabilities assigned to various CP-labelling processes (for instance $ |\mathrm{K}^0(t=0)\rangle \rightarrow |\pi\pi(t)\rangle$, $ |\overline{\mathrm{K}}^0(t=0)\rangle \rightarrow |\pi\pi\rangle$) are well-documented.

 In particular, the asymmetry $A_{\pi\pi}(t)$ has been measured by the CPLEAR experiment \cite{Angelopoulos,CPLEAR}
 %approximately
 in the range $1< t < 20~\tau_S$~.
 Figure  \ref{plot} shows two plots of the asymmetry as a function of $t$ computed in the standard WWA approach (solid curve)
 and in the T.W.F. approach (dotted curve). The left plot corresponds to the values $\tilde \epsilon_L = \epsilon \sqrt{{\Gamma_S\over \Gamma_L}}$ and $\tilde \epsilon_S = \epsilon$; the right was obtained after imposing $\tilde \epsilon_L = \epsilon \sqrt{{\Gamma_S\over \Gamma_L}}$ and $\tilde \epsilon_S = \epsilon\sqrt{{\Gamma_L\over \Gamma_s}}$ (see appendix for more explanation). In each figure, the discrepancy between the two curves is very evident
 in the range $1< t < 10~\tau_S$. Even a simple visual inspection of
 %In that region
 the CPLEAR measurement in this region (Fig.14b of Ref.~\cite{Angelopoulos}) shows that it is in perfect agreement with the WWA prediction
 and some tens of standard deviations away from the T.W.A. prediction. Therefore we can conclude that the T.W.F.
model prediction is falsified at a very high degree of confidence level.

\begin{figure}
\centering
\includegraphics[width=0.45\textwidth]{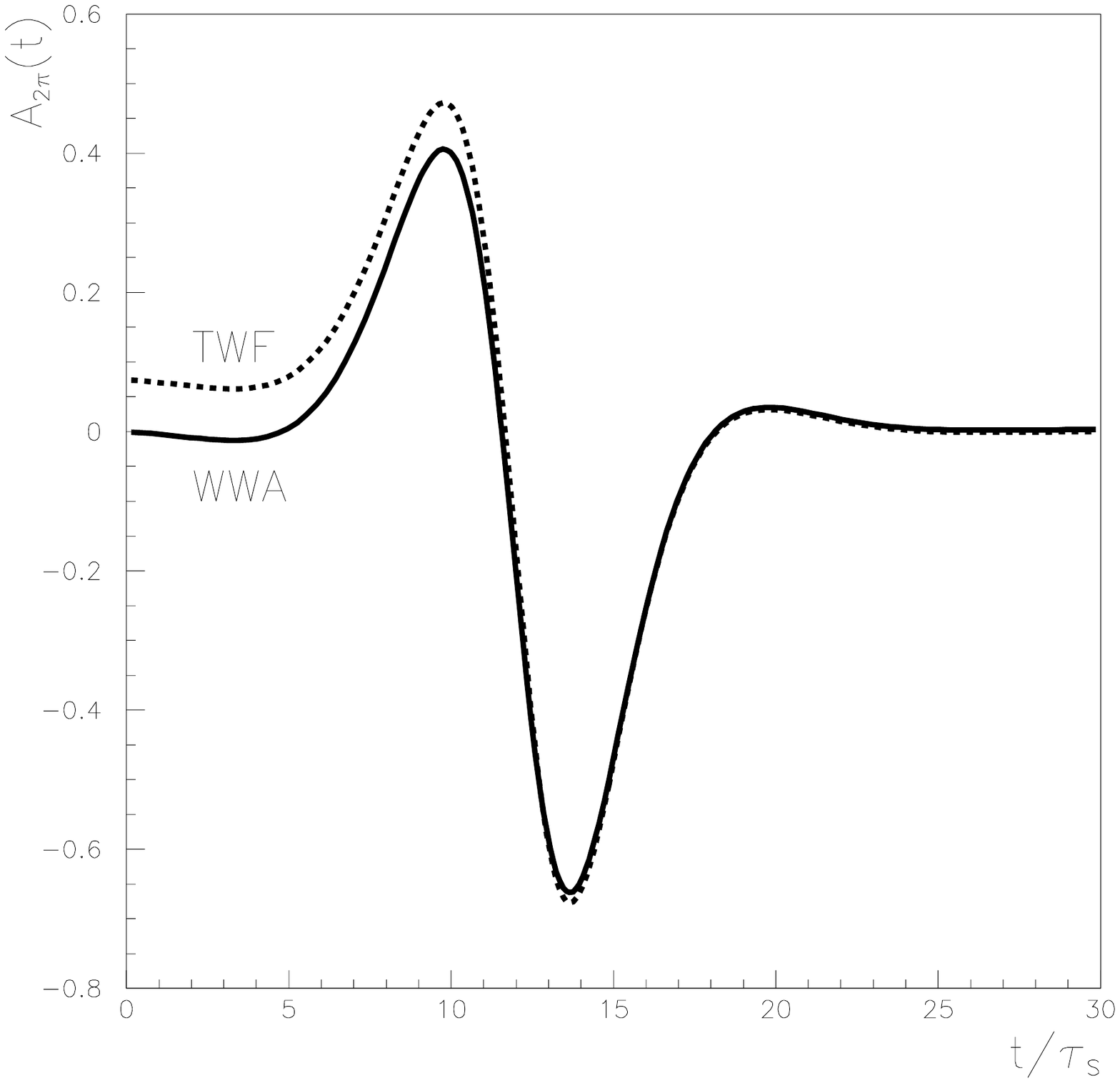}
\includegraphics[width=0.45\textwidth]{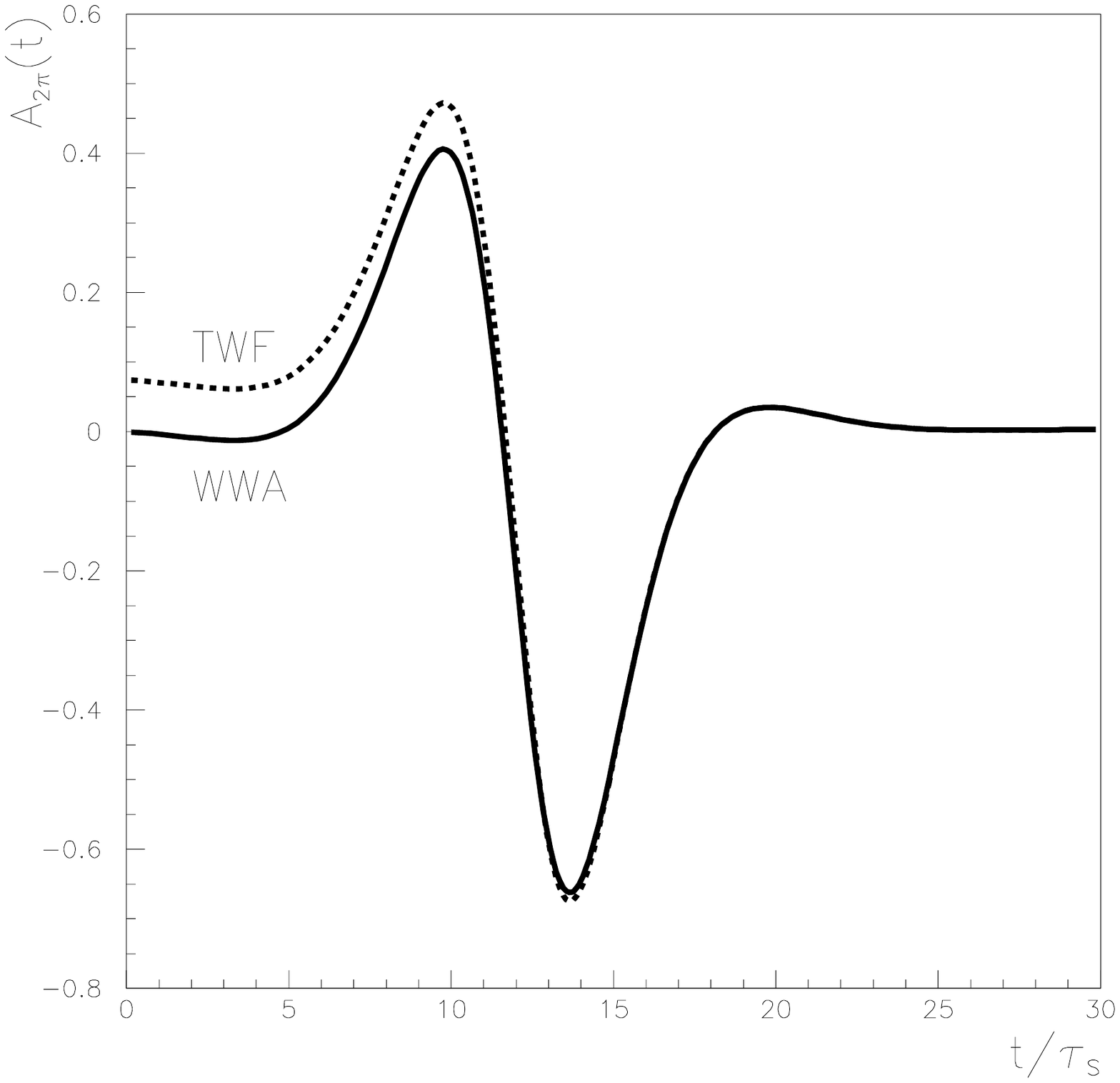}
\includegraphics[width=0.45\textwidth]{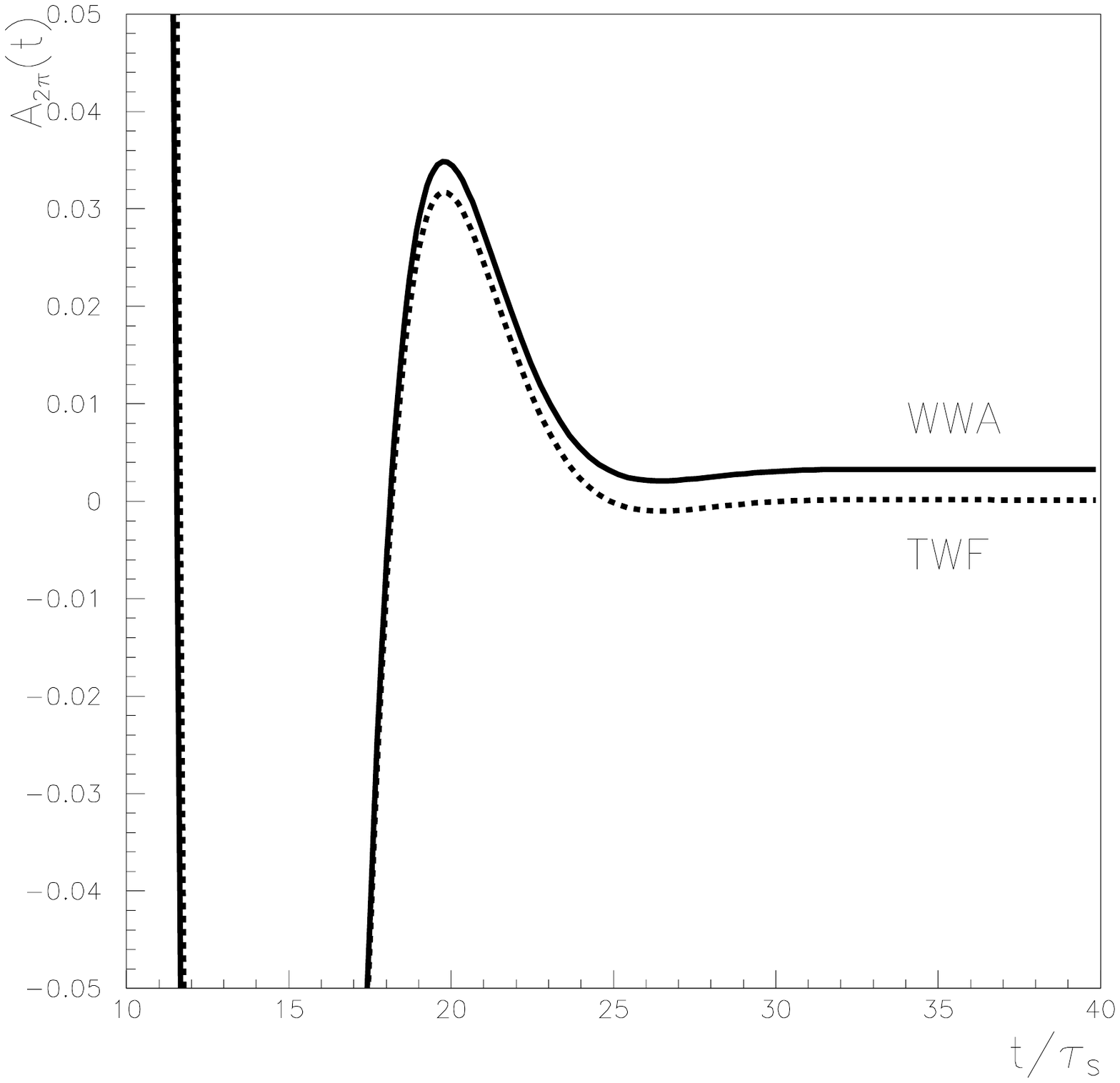}
\includegraphics[width=0.45\textwidth]{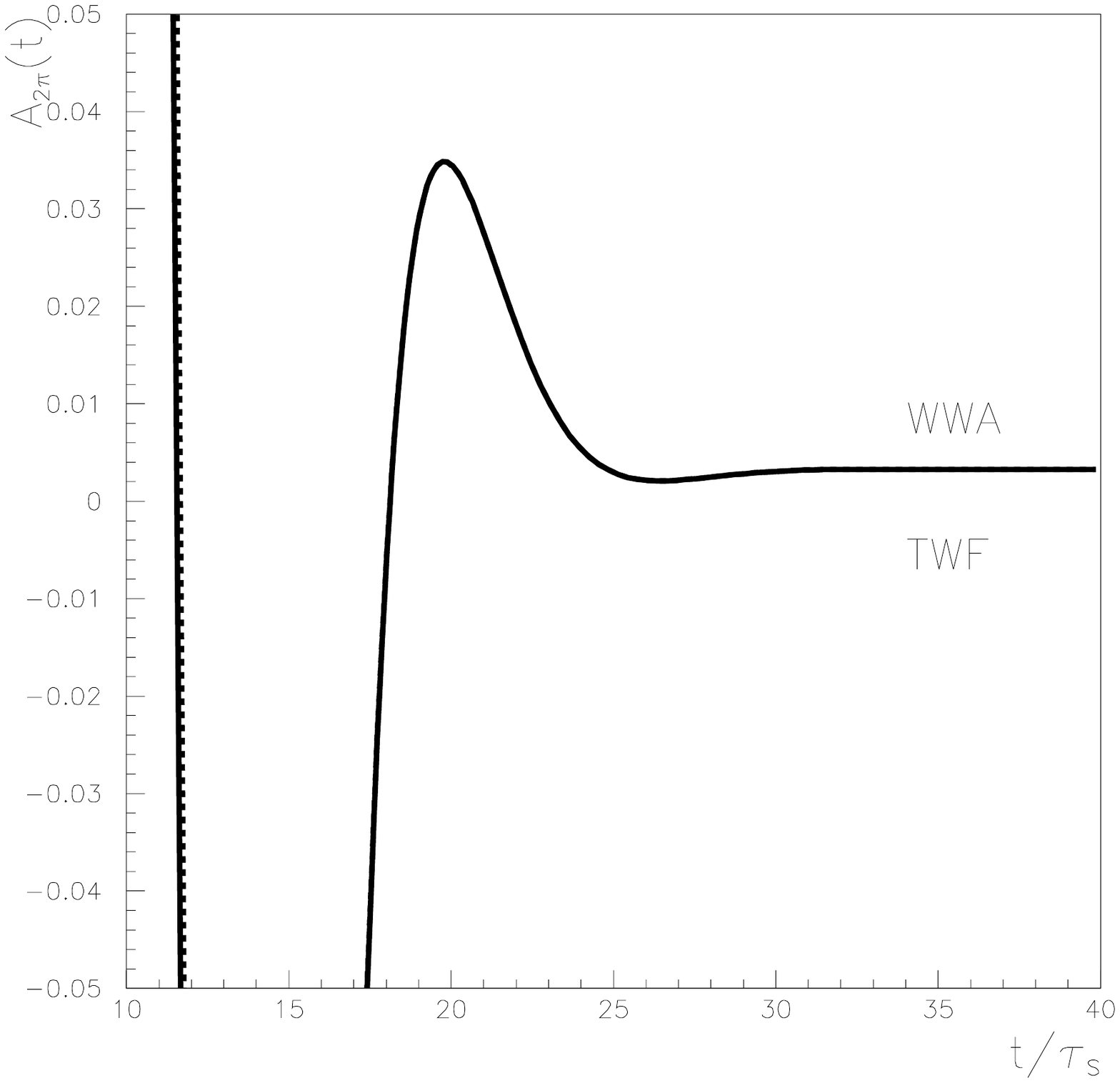}
\caption{\label{plot}
Asymmetry $A_{2\pi}(t)$ as a function of $t$ in units of $\tau_S$: standard WWA (solid curve) and T.W.F model predictions
(dotted curve). The left top plot corresponds to $\tilde \epsilon_S = \epsilon \sqrt{{\Gamma_L\over \Gamma_S}}$ and the right top plot to $\tilde \epsilon_S=\epsilon$. The significance of these values is discussed in appendix. The same plots 
with a zoomed vertical scale and in the region $10<t<40~\tau_S$ with
$\tilde \epsilon_S = \epsilon \sqrt{{\Gamma_L\over \Gamma_S}}$ (left bottom plot) and with $\tilde \epsilon_S=\epsilon$ (right bottom plot) 
show better the difference between WWA and T.W.F. model predictions in the two cases in this region.}
 \end{figure}

 % Actually, standard predictions (plain curve) were confirmed during the experiments with a precision such that for instance at $t/\tau$=4, the experimental value was definitively smaller than 0.02, even if error bars were taken into account. Comparing with figure \ref{plot}, this result constitutes a clear falsification of the T.W.F. approach.

\section{Conclusions and discussions}

Neutral kaons constitute a unique physical system where a variety of spectacular phenomena
at the root of Quantum Mechanics and Particle Physics show-up~\cite{Angelopoulos,CPLEAR,frascati, bernabeu,grimus,grimoire,frascati2,catalina}. If there were no CP-violation ($\epsilon=0$), it would be impossible to discriminate the T.W.F. model from the
standard WWA model.

In the present work, we have shown that CP violation (the discovery was awarded in 1980 with the Nobel Prize) makes instead possible to discriminate the standard approach in which Time is treated as an external, classical, parameter from the Time Operator (in particular T.W.F.) approach in which, in analogy with the spin $\frac{1}{2}$ theory, the time of occurrence of decay processes in the CP=$\pm 1$ sectors would obey a generalized Born rule (as explained in section \ref{T.W.F.section}). Indeed, as we have shown (in section \ref{falsi}), accurate experimental measurements of
neutral kaon decay rates into two pions and of the asymmetry $A_{2\pi}(t)$ showing evidence of CP violation~\cite{Angelopoulos,CPLEAR}, allow us to falsify a model based on this Time Operator approach, which constitutes the main result of our paper.

Ultimately, our analysis allows us to bring to the realm of experiments an old debate that can be traced back to the original developments of the quantum theory concerning the role and status of Time. Here CP violation turned out to be an essential ingredient of our derivation. Despite CP violation is tiny, the measurements of its manifestations are so accurate~\cite{Angelopoulos} that neutral kaons provide a successful candidate for implementing our ideas. Trapped ions also offer promising perspectives~\cite{champ}  to test other fundamental ideas concerning the status of time in the quantum theory~\cite{Massar}.

%Of course it is always possible to circumvent a no go theorem and the door remains open in principle to more sophisticated non-standard models.

 Actually our analysis is also valid when one considers other models as for instance the Time Super-Operator proposed by Misra Prigogine and Courbage~\cite{MPC2}. Here Time is a quantized variable (a q-number)~\cite{confusion}. This Time Super-Operator has been recently also applied to kaon phenomenology~\cite{TSO}. Even in that case structural discrepancies can be found and, in principle, it should be possible to discriminate these Super-Operator models from the standard theory.

%Despite of this, it is worth noting that, in last resort, nothing forbids to conceive a fully ad hoc model, along the lines sketched in section \ref{multichannel} where we associate at each decay process into final products $i$ occurring at a rate $\Gamma_i$,
%a temporal wave function of which the amplitude $\Psi^{T.W.F.}$ is proportional to $\sqrt{\Gamma_i }$ so that the standard probability at time $dt$ derived with the help of the Fermi golden rule, $|\sqrt \Gamma_i\sqrt{dt}|^2$, equals $|\sqrt \Gamma_i|^2dt=|\Psi^{T.W.F.}|^2dt$. Of course, such models are in last resort indistinguishable from the standard WWA mass-decay approach and do not present much conceptual interest precisely for this reason.

%{\it (REFERENCES BELOW SHOULD BE LISTED IN ORDER)}

\textbf{Acknowledgement:}\\
B.C.H. gratefully acknowledges the Austrian Science Fund (FWF-P26783). All of us acknowledge support from the COST MP 1006 action
(Fundamental Problems in Quantum Physics) and in particular support from the Working Group 4 – From theory to experiment.

{\section*{Appendix.}
\subsection*{Appendix 1: Fermi Golden Rule and multi-channel decay.}
Let us describe the decay process by the system of coupled equations
\begin{subequations} \label{eq:ExactEQmultiori}
  \begin{align}
    i\dot{d}_{\mathrm{in}}\left(t\right) &=\omega_{\mathrm{in}}  d_{\mathrm{in}}\left(t\right)+ \sum_{i,\lambda} G_{i,\lambda}^* \ d_{i,\lambda}\left(t\right), \label{eq:ExactEQemultiori}\\
    i\dot{d}_{i,\lambda}\left(t\right) &=  \omega_{i,\lambda}d_{i,\lambda}\left(t\right) +G_{i,\lambda} \ d_{\mathrm{in}}\left(t\right) ,\label{eq:ExactEQgmultiori}
  \end{align}
\end{subequations}
where we denoted $\omega_{\mathrm{in}}$ and $d_{in}$ the energy and amplitude of the ingoing mode (e.g. a kaon produced in a particle accelerator) while the indices $i$ and $\lambda$ are respectively assigned to different species of decay products (e.g. pion pairs and pion triplets)  and to their internal degrees of freedom (energy, direction and so on).
\\ It is convenient to introduce $c_{\mathrm{in}}=d_{\mathrm{in}}\mathrm{e}^{\mathrm{i} \omega_{\mathrm{in}} t}$ and $c_{i,\lambda}=d_{i,\lambda}\mathrm{e}^{\mathrm{i} \omega_{i,\lambda} t}$, so that we can cast (\ref{eq:ExactEQmultiori}) in the form \begin{subequations} \label{eq:ExactEQmulti}
  \begin{align}
    \dot{c}_{\mathrm{in}}\left(t\right) &= \sum_{i,\lambda} G_{i,\lambda}^* \ c_{i,\lambda}\left(t\right) \ \mathrm{e}^{-\mathrm{i} \left(\omega_{i,\lambda}-\omega_{\mathrm{in}}\right) t}, \label{eq:ExactEQemulti}\\
    \dot{c}_{i,\lambda}\left(t\right) &=   G_{i,\lambda} \ c_{\mathrm{in}}\left(t\right) \ \mathrm{e}^{\mathrm{i} \left(\omega_{i,\lambda}-\omega_{\mathrm{in}}\right)  t},\label{eq:ExactEQgmulti}
  \end{align}
\end{subequations}

In a perturbative approach, it is standard to impose that, at short time, $c_{in}(t)=c_{in}(t=0)=1$ in first approximation so that we get after a straightforward integration:
\\ $c_{i,\lambda}\left(t\right) = t\cdot \mathrm{e}^{\mathrm{i} \left(\omega_{i,\lambda}-\omega_{\mathrm{in}}\right)  t/2}G_{i,\lambda}sinc(\left(\omega_{i,\lambda}-\omega_{\mathrm{in}}\right)t/2)$, from which we derive all results of section \ref{1.3} concerning the Fermi Golden Rule.
\\ Let us now define \cite{fermi} the state
\\ $|eff.(\omega)>$ through $|eff.(\omega)>={1\over \sqrt{\sum_{i,\lambda,\omega_{i,\lambda}=\omega}|G_{i,\lambda} |^2}}\sum_{i,\lambda,\omega_{i,\lambda}=\omega}G_{i,\lambda}|i,\lambda,\omega>$.
\\ It is easy to check that, in each energy sector characterized by an energy $\omega$, all superpositions of modes orthogonal to $|eff.(\omega)>$ are decoupled from the decay process, so that in practice everything happens as if there was only one decay product, $|eff.(\omega)>$, which is a coherent superposition of different decay products, and is coupled to the ingoing mode with the effective coupling constant $G_{eff.}(\omega)=\sqrt{\sum_{i,\lambda,\omega_{i,\lambda}=\omega}|G_{i,\lambda} |^2}$.
\\ The Fermi Golden Rule treatment can also be applied in this case, and we find, following the computation sketched in section \ref{multichannel}, that the decay into this effective single mode is characterized by an effective Gamma factor $\Gamma_{eff.}$ which obeys
\\ $\Gamma_{eff.}=\sum_i\sum_\lambda D.O.S.(\omega_{in},i,\lambda) 2\pi|G_{i,\lambda}(\omega_{in})|^2=\sum_i\Gamma_i,$ where $\Gamma_i$ is the Gamma factor associated to a particular species of decay product, in agreement with section \ref{multichannel}.

\subsection*{Appendix 2: constraints on $\tilde  \epsilon_S$.}
Due to the fact that CP-violation is small, we expect that $\tilde \epsilon_S$ is a small parameter. It can be shown that if we impose that the theoretical prediction of $P_{K_S(0)\rightarrow 3\pi(t)}$, made in the standard WWA approach, is the same as the corresponding prediction, $\tilde P_{K_S(0)\rightarrow 3\pi(t)}$, made in the T.W.F. approach, we find, after computations similar to those that led to the prediction (\ref{approx}), the constraint $$  \tilde \epsilon_S = \epsilon \sqrt{{\Gamma_L\over \Gamma_S}} $$ according to which $\tilde  \epsilon_S$ is a very small parameter (of the order of 10$^{-6}$). This constraint however is not based on experimental data because no precise measure of the $3\pi$ production rate has been realized up to now~\cite{KLOE3pi0}.
\\ The measure of the asymmetry parameter provides instead another constraint on $\tilde \epsilon_S$.
\\ Indeed, it can be shown by straightforward computation that in the limit $t\gg \tau_S$ \begin{eqnarray} A_{\pi\pi}(t \gg \tau_S) &\simeq& 2 \Re(\epsilon) \simeq 3.3 \times 10^{-3} \nonumber\\
\tilde A_{\pi\pi}(t \gg \tau_S)
%\approx +2Re.((1-\sqrt{\Gamma_L\over  \Gamma_S})\cdot \tilde \epsilon_L)
%\begin{eqnarray} A_{\pi\pi}(t \gg \tau_S)
&\simeq& 2 \Re(\tilde \epsilon_S)
%\approx +2Re(\tilde \epsilon_L)\approx 8 \%\not=0\label{epsilon}
\end{eqnarray}
\\ If now we impose to get identical results in both approaches, we find
\\ $$2\Re(\tilde \epsilon_S) = 2 \Re(\epsilon )\simeq 3.3 \times 10^{-3},$$ which is satisfied if $\epsilon=\tilde \epsilon_S$.
\\ Obviously, the above result would not be true in the case $\tilde \epsilon_S \neq \epsilon$.
\\ The two predictions made above ($\tilde  \epsilon_S$=$\epsilon \sqrt{{\Gamma_L\over \Gamma_S}}$ and $\tilde \epsilon_S = \epsilon$) are plotted in figure 1. Without zooming (bottom plots), the difference between the two top plots is nearly imperceptible with the naked eye, but this does not really matter: what is important is the constraints derived from observations on $\tilde \epsilon_L$ alone, and those constraints suffice to discriminate the T.W.F. approach as we have shown in the paper.

%\begin{eqnarray}lim_{t\rightarrow 0} A^{CPT-violating}_{\pi\pi}(t)=\nonumber\\
%lim_{t\rightarrow 0}
%{({\Gamma_{K_1\rightarrow \pi\pi}\over 2|1- \epsilon_S \epsilon_L|^2})(|\sqrt{1+|\epsilon_S|^2}(1- \epsilon_L)exp^{-iE_St/\hbar}+\sqrt{1+|\epsilon_L|^2}(1-\epsilon_S) \epsilon_Lexp^{-iE_Lt/\hbar}|^2)\over 2 ({\Gamma_%{K_1\rightarrow \pi\pi}\over 2|1-\tilde \epsilon_S\tilde \epsilon_L|^2})}
%\nonumber \\
%-lim_{t\rightarrow 0}
%{({\Gamma_{K_1\rightarrow \pi\pi}\over 2|1-\tilde \epsilon_S\tilde \epsilon_L|^2})(\sqrt{1+|\epsilon_S|^2}|(1+ \epsilon_L)exp^{-iE_St/\hbar}-\sqrt{1+|\epsilon_L|^2}(1+ \epsilon_S) \epsilon_L exp^{-iE_Lt/\hbar}|^2)\over 2({\Gamma_{K_1\rightarrow \pi\pi}\over 2|1- \epsilon_S \epsilon_L|^2})},\label{epsilonviol}\end{eqnarray}

%where $\epsilon_L$ and $\epsilon_R$ are small parameters for which it is known from kaon phenomenology that $\epsilon_L\approx \epsilon$ and $|\epsilon_S|<0.009$ at 90 $\%$ C.L..

%It is indeed easy to show that $lim_{t\rightarrow 0} A^{CPT-violating}_{\pi\pi}(t)$ contains only quartic contributions in $\epsilon_L$ and $\epsilon_R$ so that

%\begin{equation}lim_{t\rightarrow 0} A^{CPT-violating}_{\pi\pi}(t)\approx 0.\end{equation}
%{\\ \it ===================== \\ }

 \end{document}